\documentclass[pdflatex,sn-mathphys-num]{sn-jnl}

\usepackage{graphicx}%
\usepackage{multirow}%
\usepackage{amsmath,amssymb,amsfonts}%
\usepackage{amsthm}%
\usepackage{mathrsfs}%
\usepackage[title]{appendix}%
\usepackage{xcolor}%
\usepackage{textcomp}%
\usepackage{manyfoot}%
\usepackage{booktabs}%
\usepackage{algorithm}%
\usepackage{algorithmicx}%
\usepackage{algpseudocode}%
\usepackage{listings}%

\usepackage{colortbl}
\usepackage{tcolorbox}
\usepackage{color}
\usepackage{multirow}
\usepackage{makecell}
\usepackage{tabularx}
\usepackage{graphicx}
\usepackage{soul}
\usepackage{tikz}
\usepackage[section]{placeins}
\usepackage{comment}
\usepackage{booktabs}
\usepackage{verbdef}
\usepackage{pifont}           
\usepackage{balance}
\usepackage[]{hyperref}
\usepackage{url} 
\usepackage{lmodern}
\usepackage{anyfontsize}
\tcbuselibrary{skins} 

\usepackage{todonotes}
\setuptodonotes{fancyline, color=blue!30, size=\footnotesize}

\hypersetup{                   
  colorlinks,
  linkcolor={green!80!black},
  citecolor={red!70!black},
  urlcolor={blue!70!black}
}

\def\BibTeX{{\rm B\kern-.05em{\sc i\kern-.025em b}\kern-.08em
    T\kern-.1667em\lower.7ex\hbox{E}\kern-.125emX}}

\definecolor{customgrey}{RGB}{223, 223, 223}
\newcommand{\toolname}[1]{\emph{FuzzBox}}
\newcommand*\circled[1]{\tikz[baseline=(char.base)]{
            \node[shape=circle,draw,inner sep=1pt,font=\sffamily\footnotesize] (char) {\textbf{#1}};}} 

\newcommand{\xmark}{\ding{55}}%

\NewTotalTColorBox[auto counter]{\Definition}{ +m }{ 
    notitle,
    colback=blue!3!white,
    frame hidden,
    boxrule=0pt,
    enhanced,
    sharp corners,
    borderline west={4pt}{0pt}{blue!20!white},
     boxsep=2pt,  
    left=7pt,    
    right=2pt,   
    top=2pt,     
    bottom=2pt,  
}{
    \textcolor{green!50!black}{
        \sffamily
    }
    #1
}

\begin{document}

\title[FuzzBox: Blending Fuzzing into Emulation \\for Binary-Only Embedded Targets]{FuzzBox: Blending Fuzzing into Emulation for Binary-Only Embedded Targets}

\author{\fnm{Carmine} \sur{Cesarano}}\email{\{carmine.cesarano2, roberto.natella\}@unina.it}

\author{\fnm{Roberto} \sur{Natella}}

\affil{\orgdiv{DIETI}, \orgname{Università degli Studi di Napoli Federico II}, \orgaddress{\city{Naples}, \state{Italy}}}

\abstract{Coverage-guided fuzzing has been widely applied to address zero-day vulnerabilities in general-purpose software and operating systems. This approach relies on instrumenting the target code at compile time. However, applying it to industrial systems remains challenging, due to proprietary and closed-source compiler toolchains and lack of access to source code. \toolname{} addresses these limitations by integrating emulation with fuzzing: it dynamically instruments code during execution in a virtualized environment, for the injection of fuzz inputs, failure detection, and coverage analysis, without requiring source code recompilation and hardware-specific dependencies. We show the effectiveness of \toolname{} through experiments in the context of a proprietary MILS (Multiple Independent Levels of Security) hypervisor for industrial applications. Additionally, we analyze the applicability of \toolname{} across commercial IoT firmware, showcasing its broad portability.}


\maketitle

\section{Introduction}\label{sec1}
\label{sec:introduction}
In the realm of software security, the persistent threat of zero-day vulnerabilities poses a significant challenge to maintaining the safety and reliability of software systems. Fuzzing has proven to be one of the most effective techniques for discovering such vulnerabilities, particularly in memory management, by generating a large volume of fuzz inputs through mutations \cite{zeller2019fuzzing, manes2018fuzzing, eisele2022embedded}. This is typically achieved by linking the target software with a client (\emph{fuzz driver}) that repeatedly applies fuzz inputs and detects whether the target experienced a failure (e.g., a crash or timeout). To achieve better effectiveness, \emph{grey-box} fuzzing also collects information about code coverage and adds a feedback loop to keep and mutate only inputs that reach new code coverage, in order to steer the fuzzing toward unexplored parts of the target. 

While fuzzing has been widely applied to general-purpose programs and operating systems, extending its applicability to industrial embedded systems remains a challenging task \cite{yun2022fuzzing, eisele2022embedded}. Existing fuzzers assume that the target system can be modified to introduce a fuzz driver, and instrumented with probes to obtain feedback about failure events and code coverage. While this is feasible for general-purpose systems using toolchains like LLVM \cite{llvm} or GCC \cite{gcc} by adding new monitoring instructions in the compiled binary, industrial embedded systems often rely on proprietary or outdated toolchains that lack support for code instrumentation. Existing solutions that perform static instrumentation of compiled binaries (binary rewriting) \cite{afl-dyninst,dinesh2020retrowrite} are also not applicable, since they are limited to user-space Linux binaries and are not portable across different architectures and operating systems. Other approaches \cite{AFL-DynamoRIO,AFL-PIN} instrument the compiled binary at run-time (dynamic binary instrumentation), but they are also limited to user-space Linux binaries, and they incur a significant performance overhead. Finally, some approaches make use of hardware features, such as Intel PT \cite{chen2019ptrix, schumilo2017kafl, honggfuzz} and debugging interfaces \cite{eisele2023fuzzing, li2022muafl}. However, this approach is dependent on specific hardware, limiting portability.

In this paper, we present \toolname{}, a novel fuzzing tool designed to integrate fuzzing with emulation, addressing key challenges in industrial embedded systems, in particular those based on proprietary operating systems and toolchains. Unlike existing approaches that rely on static or dynamic binary instrumentation, \toolname{} employs QEMU-based system-level emulation to transparently intercept target functions, modify the guest's state to inject fuzz inputs into the target, and monitor its execution for detecting failures and for tracking coverage. This approach eliminates the need to rebuild the target and to introduce a custom fuzz driver, and it enables fuzzing across diverse CPU architectures and software configurations. The proposed solution fits well the constraints of industrial embedded systems, which typically adopt proprietary compiler toolchains, heterogeneous hardware platforms, and diverse software configurations.

In our experimental analysis, we show the benefits of \toolname{} in the context of a proprietary embedded hypervisor (\emph{Wind River VxWorks MILS} \cite{neugass1991vxworks}), which is adopted as a separation kernel in many security-critical industrial domains such as automotive, railways, avionics, and defense. Furthermore, we conducted a fuzzing campaign on Linux-based IoT firmware, to show the portability of our tool across diverse embedded targets. On average, our approach achieved a 2X increase in fuzzing throughput, a coverage improvement of approximately 58\%, and faster bug finding compared to black-box fuzzing.

In summary, the paper presents the following contributions:
\begin{itemize}
    \item The design of \toolname{}, a fuzzing approach that integrates fuzzing with emulation to address challenges in industrial embedded systems.
    \item The implementation of the tool based on QEMU, enabling support for a wide range of CPU architectures.
    \item An experimental evaluation of \toolname{} in terms of fuzzing throughput, coverage, and bugs found.
\end{itemize}

We publicly released \toolname{}\footnote{\url{https://github.com/dessertlab/FuzzBox}}. The remainder of the paper is organized as follows. Section \ref{sec:background} provides a motivating example and technical background. Section \ref{related_work} discusses related work. Sections \ref{sec:design} and \ref{sec:implementation} present the design and implementation of \toolname{}. Sections \ref{sec:evaluation} and \ref{sec:iot} provide an experimental evaluation of \toolname{}. Section \ref{sec:discussion} discusses limitations. Section \ref{sec:conclusion} concludes the paper.

\section{Background and Motivation}
\label{sec:background}
In this section, we provide background and a motivating example to highlight the challenges in fuzzing industrial embedded systems, in the context of the VxWorks MILS hypervisor as a representative case.

\subsection{MILS}
\label{sec:MILS_gateway}
MILS (Multiple Independent Levels of Security) \cite{alves2006mils} is a high-assurance security architecture, based on the concept of separation and controlled information flow. MILS products, including OSes and hypervisors such as Wind River VxWorks MILS, act as a \emph{separation kernel}, which allows multiple applications with varying security levels to coexist on the same hardware platform. These kernels ensure strict isolation, disciplined communication, and resource access control.

\begin{figure}[t]
  \centering
  \includegraphics[width=0.7\linewidth]{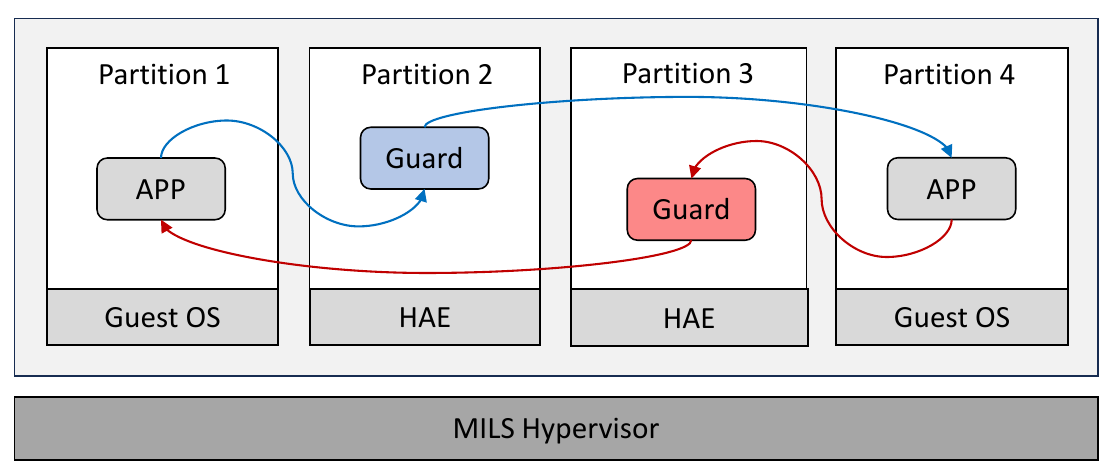}
  \caption{Multiple Independent Levels of Security (MILS) architecture}
  \label{fig:MILS_architecture}
\end{figure}

A relevant use case for MILS technology is the \emph{MILS-based gateway} \cite{netkachova2015security}, which supports secure communication beetween distinct network domains with differing security classifications. The gateway routes and filters data between domains, e.g., blocking data leaks from a ``\textit{sensitive}'' network to the ``\textit{unrestricted}'' network \cite{airlines2005arinc}. This use case can be implemented in VxWorks MILS (\figurename{}~\ref{fig:MILS_architecture}) with a set of separated \emph{virtual boards} (i.e., virtual machines, with strict real-time constraints). For example, two virtual boards are assigned by the kernel to two Ethernet interfaces and interact with the two separate network domains, acting as network proxies; and two additional virtual boards (i.e., ``\textit{guards}'') that analyze and filter the traffic from each network domain. Thus,  traffic flows from one network domain through the four virtual boards and then to the other network domain. Secure communication between virtual boards is implemented using Secure Inter-Partition Communication channels (SIPC) \cite{yang2009inter}. Additional virtual boards may handle system administration, logging, and auditing, and other additional functions (e.g., self-testing).

The MILS architecture can support many diverse and complex system configurations. Therefore, MILS products provide proprietary toolchains to build a system. For example, the VxWorks Workbench \cite{windriver_workbench} uses an XML-based configuration to define system parameters such as physical board types, reference processor, memory layout, device mapping, SIPC communication channels between partitions, and security policies on allowed traffic. The workbench allows developers to configure the criticality of each partition, and the type of kernel layer to use. In the MILS-based gateway use case (Figure \ref{fig:MILS_architecture}), the setup uses VxWorks Guest OS, a Linux-based kernel, for low-criticality partitions; and the High Assurance Environment (HAE), a minimal run-time library, for high-criticality partitions. The workbench facilitates the development of user-space applications, executed as periodic tasks on virtual boards. Finally, the workbench builds the MILS hypervisor, guest OSes, and user-space application into a single binary. This binary can then be loaded onto a physical board for execution.

\subsection{Motivating Example}
\label{sec:motivating_example}
Fuzzing industrial embedded systems applications poses unique challenges. In this section, we discuss these challenges in the context of MILS-based systems, motivating the design of \toolname{}.

\bigbreak
\noindent
\textbf{\textit{Intrusiveness.}} 
In order to perform fuzzing, it is necessary to introduce a \emph{fuzz driver} \cite{ispoglou2020fuzzgen} that submits fuzz inputs to the target software. In the case of MILS-based applications, the target software runs within a virtual board (e.g., in the case of a MILS-based gateway, the ``\textit{guard}'' virtual board), and interacts with other components through SIPC channels (which represent the attack surface of the target), in a client-server fashion. Therefore, fuzzing a MILS-based application requires the introduction of a fuzz driver in a dedicated virtual board, which acts as a client to send fuzz inputs. This modification requires a significant engineering effort since the target system needs to be rewritten, reconfigured, and rebuilt. Moreover, this may be difficult or not feasible if no source code is available (e.g. when auditing proprietary binary code). 

Another challenge is represented by the lack of visibility on the execution of the target application. Fuzzing requires the fuzz driver to detect the occurrence of a target failure. However, this may be difficult to achieve in special-purpose embedded systems. For example, in the case of the MILS kernel, there is no software interface to determine the failure of a virtual board from another virtual board. A common approach is to detect failures from the client, using liveness checks \cite{feng2021snipuzz} to periodically probe the state of the target application. However, the application may not be designed to send replies to the clients, such as in the case of the MILS-based gateway. If responses are unavailable, an additional SIPC channel must be implemented, requiring customization of both the client and server to facilitate the exchange of liveness checks over SIPC. This modification requires significant engineering effort, since it changes the architecture of the target software. Furthermore, the use of response messages as an indication of failures can be prone to false alarms. For instance, the target application might exhibit delayed response times, especially when handling unusually large fuzz input or due to non-deterministic factors (e.g., external events, concurrency). In these cases, the fuzz driver can mistakenly conclude that the target application failed.

\bigbreak
\noindent
\textbf{\textit{Toolchain Dependency.}} 
Even if a developer is willing to invest engineering efforts to implement a fuzz driver and additional mechanisms for failure detection, industrial embedded systems do not provide sophisticated solutions for coverage-driven (i.e., grey-box) fuzzing. Coverage-driven fuzzing improves effectiveness by tracking which code blocks are executed during fuzzing, in order to steer testing towards unexplored parts of the target. However, obtaining coverage information is not readily accessible to the fuzz driver. Proprietary OSes and compilers for embedded systems do not provide advanced features for code instrumentation, differently from tools for general-purpose systems such as LLVM and GCC. Moreover, code coverage in embedded systems is typically not designed for real-time use by a fuzzer, but it is only limited to offline analysis (e.g., for reporting purposes, similarly to \textit{gcov} \cite{gcc}). Therefore, coverage information is not available to applications running in the system, but only to external tools (e.g., debuggers). In the specific case of the MILS-based gateway, the entire project, including all partitions, must be compiled, linked, and packaged with the proprietary \emph{VxWorks Workbench} toolchain. Workbench neither supports code instrumentation (such as the option \texttt{-fsanitize=coverage} provided by open-source compilers), nor exposes plug-in hooks (such as Clang and GCC plug-ins), which forces fuzzing to operate in blind black-box mode. Even with source code available, inserting custom passes for code instrumentation would require to write new tools from scratch for that specific target.

\bigbreak
\noindent
\textbf{\textit{Hardware Dependency.}} 
Typically, embedded systems do not adopt CPU architectures for general-purpose systems (e.g., x86), but they are based on specialized CPU architectures such as ARM, MIPS, PowerPC and SPARC. For example, the Wind River VxWorks MILS comes with a board support package for the PowerPC architecture. Consequently, this diversity poses a significant challenge for embedded system fuzzing, demanding support for multiple architectures. In addition, these architectures lack advanced hardware features that modern fuzzers use. When compiler-assisted instrumentation cannot be enabled, as in the case for the VxWorks MILS, state-of-the-art fuzzers typically fall back on hardware tracing units (e.g., Intel PT) \cite{chen2019ptrix} or debugger-assisted tracing via GDB \cite{eisele2023fuzzing} to recover execution feedback. Unfortunately, MILS-based boards \cite{vxworks_board} do not offer neither Intel PT nor a GDB Remote Serial Protocol stub, leaving fuzzers without any practical source of execution traces.

\bigbreak
In conclusion, the heterogeneity of embedded devices, the need for modifying the target to insert fuzz drivers, and the challenges in recompiling, instrumenting, or leveraging hardware support are obstacles for effective fuzzing of embedded systems such as MILS-based applications. The proprietary Workbench toolchain and PowerPC‑based hardware of our motivating gateway exemplify how toolchain and hardware dependencies jointly deprive fuzzers of compiler‑level, debugger‑level, and hardware‑assisted feedback. Addressing these challenges is crucial for effective fuzzing.

\section{Related Work}
\label{related_work}
Existing fuzzing techniques have shown high effectiveness at uncovering vulnerabilities in general-purpose software and operating systems. However, their application to industrial embedded systems, such MILS-based applications, faces substantial limitations due to the unique challenges of these systems. First, the target is often available only as a binary, so the fuzzer must cope without source code (\textbf{C1~Binary-only}). Second, certification constraints forbid patching the executable or inserting in-system fuzz agents, calling for a strictly non-intrusive test harness (\textbf{C2~Non-intrusive}). Third, proprietary build pipelines rarely expose compiler instrumentation, so any practical solution should avoid reliance on advanced tool-chains (\textbf{C3~Tool-chain independence}). Fourth, many embedded boards lack Intel PT, or even standard GDB stubs, demanding an approach that remains agnostic to hardware tracing features (\textbf{C4~Hardware independence}). Finally, security assurance requires the ability to drive inputs through every layer of the stack—from user processes to kernels and hypervisors—hence the need for full-system coverage with arbitrary entry points (\textbf{C5~Full-system scope}). In the remainder of this section we review the main lines of prior work and show their limitations in addressing the challenges of industrial embedded systems.

\subsection{Static Binary Rewriting}
Fuzzing techniques based on static binary rewriting techniques, such as AFL-Dyninst~\cite{afl-dyninst}, RetroWrite~\cite{dinesh2020retrowrite} and AflIoT~\cite{du2022afliot}, instrument binaries without requiring source code by analyzing and modifying their compiled structure. These approaches use framework such as Dyninst to inject instrumentation directly into the binary, enabling fuzzing by tracking code coverage. However, current rewriters are implemented only for mainstream ISAs such as x86 and AArch64, leaving PowerPC, SPARC, or other architectures typical of industrial platforms unsupported and therefore falling short on \textbf{C4~(Hardware independence)}.  Finally, static rewriting has so far proved practical only for user-space programs; extending it to kernels, hypervisors, or the multi-partition layout of a MILS system remains an open challenge, meaning \textbf{C5~(Full-system scope)} is unmet.

\subsection{Dynamic Binary Instrumentation}
Approaches to fuzzing which use dynamic binary instrumentation (DBI), such as AFL-DynamoRIO~\cite{AFL-DynamoRIO} and AFL-PIN~\cite{AFL-PIN}, insert probes at runtime via frameworks such as DynamoRIO and Intel PIN. Because the probes are injected into an already-compiled executable, the technique works with opaque binaries. It also operates without compile-time support and thus satisfies tool-chain independence. Their portability, however, is bounded by the underlying DBI framework: DynamoRIO supports multiple ISAs yet still omits several industrial targets, and Intel PIN is restricted to x86/AMD64.  Consequently, DBI cannot be regarded as fully \textbf{C4~(Hardware independence)} when the deployment board is PowerPC, SPARC, or another non-Intel architecture typical of MILS gateways. Moreover, mainstream DBI frameworks operate only at user privilege level, leaving kernel and hypervisor code out of scope, and therefore fall short of \textbf{C5~(Full-system scope)} for industrial embedded deployments.

\subsection{Hardware-Assisted Techniques}
Hardware-assisted fuzzing techniques, including Ptrix~\cite{chen2019ptrix}, Honggfuzz~\cite{honggfuzz}, kAFL~\cite{schumilo2017kafl}, $\mu$AFL~\cite{li2022muafl}, and GDBFuzz~\cite{eisele2023fuzzing}, leverage features such as Intel Processor Trace (Intel PT), debugging interfaces, or other hardware-assisted tracing capabilities to collect coverage information and monitor execution. These techniques can achieve high precision and efficiency in specific environments working on opaque binaries and requiring neither compile-time instrumentation nor intrusive modifications. However, their limitations is tied to \textbf{C4~(Hardware independency)}: Intel PT and x86-centric debug stubs are absent on the PowerPC, MIPS and ARM boards, that dominate MILS and other industrial deployments, limiting the applicability of this approach in these domains. Only a subset of these tools (e.g., GDBFuzz and kAFL) extends tracing into kernel mode, so full coverage of the entire stack is not guaranteed, and \textbf{C5~(Full-system scope)} is met only partially. Similarly, AflIoT~\cite{du2022afliot} introduces a hardware-dependent approach to fuzzing by executing directly on IoT devices rather than relying on emulation. While this enables direct interaction with peripheral hardware, it also restricts AflIoT to devices that support its instrumentation method, limiting its applicability to broader embedded system architectures that require platform-agnostic fuzzing solutions.

\subsection{Emulation-Based Approaches}
Emulation-based user-space fuzzers, such as AFL’s QEMU mode~\cite{afl}, AFL++~\cite{fioraldi2020afl++}, and UnicoreFuzz~\cite{maier2019unicorefuzz} rely on QEMU user-mode to run the binary file (e.g., a binary in ELF format) of the application to be fuzzed. Such binaries are meant to be executed atop a general-purpose host OS kernel, such as Linux. QEMU user-mode emulation runs these binaries by translating their CPU instructions to the host architecture, and by intercepting system calls, which are then executed via the host OS kernel. The emulation layer enables dynamic binary instrumentation of userspace program, supporting gray-box fuzzing without source code access or compile-time instrumentation. However, such tools fall short of addressing \textbf{C5~(Full-system scope)}. In fact, these tools cannot run a full embedded system binary, such as in the case of the VxWorks MILS gateway, which includes the RTOS, hypervisor, partitions, and application code all linked into a single binary image that expects to run bare metal. Even in cases where industrial components are shipped as standalone user-space binaries, they often depend on proprietary real-time OSes with non-Linux syscall ABIs and custom runtime loaders, making QEMU user-mode emulation incompatible.

\subsection{Fuzzing for Specialized Targets}
Tools like Syzkaller~\cite{vyukov2015syzkaller}, Tardis~\cite{shen2022tardis} and AflIoT~\cite{du2022afliot} are specifically designed for fuzzing operating system kernels, embedded software and IoT binaries. Syzkaller and Tardis focus on kernel fuzzing, leveraging custom fuzz drivers and source-level instrumentation to generate inputs and monitor execution, so they fail \textbf{C1 (Binary-only)}, \textbf{C2 (Non-Intrusive)} and \textbf{C3 (Tool-chain independence)}. While they achieve deep coverage of the kernel, they leave user partitions and hypervisors untouched, falling short of \textbf{C5 (Full-system scope)}. AflIoT, in contrast, is tailored for on-device fuzzing of Linux-based IoT binaries, leveraging static binary instrumentation to enable fuzzing directly on IoT devices instead of relying on emulation. While these tools are highly effective for their intended targets, they require specific runtime environments or significant modifications to the build process, making them unsuitable for full-system binary-only proprietary systems like industrial embedded software and MILS-based applications.

\bigbreak
Table~\ref{tab:comparative_analysis} summarizes the limitations of state-of-the-art binary fuzzing tools when applied to specialized industrial systems, such as MILS-based applications. A key limitation is the inability to support full-system fuzzing across the entire software stack, targeting any entry-point for fuzzing, including user-space libraries, OS kernels, and hypervisors. \toolname{} addresses these gaps with a non-intrusive, emulation-based approach that leverages QEMU-based system-level emulation. It enables binary-only fuzzing without requiring source code, instrumentation, or modifications to the target, ensuring independence from proprietary toolchains and compatibility with diverse platforms. 

\begin{table}[h]
\centering
\scriptsize
\caption{Limitations of the State-of-the-Art tools for binary-level fuzzing.}
\label{tab:comparative_analysis}
\begin{tabularx}{\columnwidth}{lccccc}
\toprule
\textbf{Technique} & \textbf{Binary} & \textbf{Non-} & \textbf{Tool} & \textbf{Hw} & \textbf{Full-} \\
 & \textbf{Fuzzing} & \textbf{Intrusive} & \textbf{Independ.} & \textbf{Independ.} & \textbf{System} \\
\midrule

AFL-Dyninst~\cite{afl-dyninst}          & \checkmark & \checkmark & \checkmark & \xmark & \xmark \\
RetroWrite~\cite{dinesh2020retrowrite}  & \checkmark & \checkmark & \checkmark & \xmark & \xmark \\
AflIoT~\cite{du2022afliot}              & \checkmark & \xmark & \checkmark & \xmark & \xmark \\
 \arrayrulecolor{customgrey}\hline\arrayrulecolor{black}
AFL-DynamoRIO~\cite{AFL-DynamoRIO}      & \checkmark & \checkmark & \checkmark & \xmark & \xmark \\
AFL-PIN~\cite{AFL-PIN}                  & \checkmark & \checkmark & \checkmark & \xmark & \xmark \\
 \arrayrulecolor{customgrey}\hline\arrayrulecolor{black}
Ptrix~\cite{chen2019ptrix}              & \checkmark & \checkmark & \checkmark & \xmark & \xmark \\
Honggfuzz~\cite{honggfuzz}              & \checkmark & \checkmark & \checkmark & \xmark & \xmark \\
kAFL~\cite{schumilo2017kafl}            & \checkmark & \xmark & \checkmark & \xmark & \checkmark \\
$\mu$AFL~\cite{li2022muafl}             & \checkmark & \checkmark & \checkmark & \xmark & \xmark \\
GDBFuzz~\cite{eisele2023fuzzing}        & \checkmark & \checkmark & \checkmark & \xmark & \checkmark \\
 \arrayrulecolor{customgrey}\hline\arrayrulecolor{black}
AFL QEMU mode~\cite{afl}                & \checkmark & \checkmark & \checkmark & \checkmark & \xmark  \\
AFL++ QEMU mode~\cite{fioraldi2020afl++}& \checkmark & \checkmark & \checkmark & \checkmark & \xmark  \\ 
UnicoreFuzz~\cite{maier2019unicorefuzz} & \checkmark & \checkmark & \checkmark & \checkmark & \xmark  \\
\arrayrulecolor{customgrey}\hline\arrayrulecolor{black}
Syzkaller~\cite{vyukov2015syzkaller}    & \xmark & \xmark & \xmark & \checkmark & \xmark \\
Tardis~\cite{shen2022tardis}            & \xmark & \xmark & \xmark & \checkmark & \xmark \\
\arrayrulecolor{customgrey}\hline\arrayrulecolor{black}

\textbf{\toolname{}} & \textbf{\checkmark} & \textbf{\checkmark} & \textbf{\checkmark} & \textbf{\checkmark} & \textbf{\checkmark} \\

\bottomrule
\end{tabularx}
\end{table}

\section{\toolname{} Design}
\begin{figure}[h]
  \centering
  \includegraphics[width=0.9\linewidth]{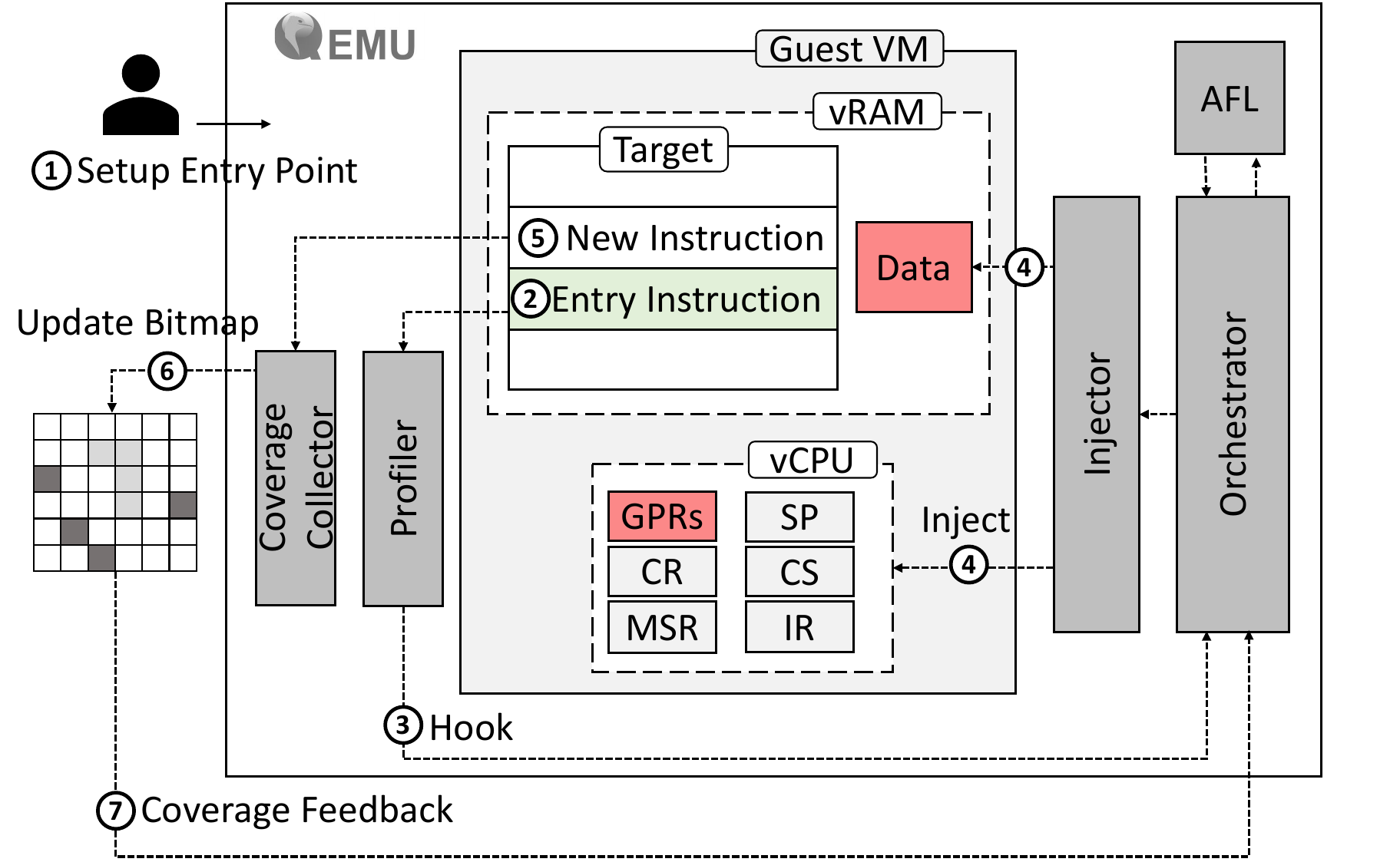}
  \caption{An Overview of the Proposed Approach \toolname{}}
  \label{fig:fuzzer_workflow}
\end{figure}

\label{sec:design}
The design of \toolname{} is based on the following design principles, in order to address the challenges discussed in the previous section:

\begin{enumerate}
    \item \textbf{Non-Intrusive}: The tool should avoid any modification of the target software, such as the introduction of a fuzz driver to trigger the target.  
    \item \textbf{Toolchain independency}: The tool should avoid depending on the availability of build toolchains able to perform code instrumentation. It should still be able to collect information about code coverage and should support binary-only (e.g., closed-source) targets.
    \item \textbf{Hardware independency}: The tool should avoid any reliance on hardware features, in order to support a broad applicability across different embedded systems.
\end{enumerate}

Figure \ref{fig:fuzzer_workflow} shows an overview of the workflow of \toolname{}. The architecture is flexible and configurable, supporting various operations for fuzzing. Users initiate the process by specifying a \emph{configuration} for fuzzing, such as the entry points of the target to be fuzzed (step \circled{1} in Figure \ref{fig:fuzzer_workflow}). Following this, \toolname{} employs the QEMU emulator \cite{bellard2005qemu} to execute the unmodified binary of the target software within a guest VM in full system mode. Leveraging QEMU's introspection capabilities, \toolname{} intercepts and fuzzes configured target functions by profiling and manipulating the guest VM state (steps \circled{2}, \circled{3}, \circled{4} in Figure \ref{fig:fuzzer_workflow}). While executing fuzzed operations, \toolname{} actively monitors the guest for abnormal behavior. Furthermore, for effective input generation, \toolname{} transparently gathers coverage information within the QEMU emulation engine (steps \circled{5}, \circled{6} in Figure \ref{fig:fuzzer_workflow}). 
This design blends fuzzing into the emulation environment, which allows us to address challenges in fuzzing industrial systems, and enables the applicability of \toolname{} to a broad range of embedded systems based on different CPU architectures and hardware platforms.

It is worth noting that \toolname{} enables fuzzing binary-only industrial firmware, including transparent instrumentation, and crash detection. It does not propose new strategies for generating fuzz inputs (e.g., input-mutation and scheduling heuristics), as these aspects are orthogonal to our proposed solution. In our implementation, we integrate the state-of-the-art \textit{LibAFL} library into \toolname{}, in order to align with the recent advances in fuzz input generation strategies, and to ensure compatibility with future developments in this area.

\subsection{Interception}
\label{sec:interception}
The \toolname{} fuzzing process is designed to intercept the invocation of a configured \textit{target function} during the normal execution of the target software and inject a new fuzz input. In this context, a \textit{fuzz input} refers to the specific values injected as input parameters of the target function, affecting the execution path of the target and potentially discovering vulnerabilities. The interception capability is carried out by two components in \toolname{}, the \textit{Profiler} and the \textit{Injector}.

\bigbreak
\noindent
\textbf{\textit{Profiler.}} 
It profiles the execution of the target software, according to the configuration of events to track. The profiler seamlessly integrates into the QEMU translation process. QEMU includes Tiny Code Generator (TCG), which translates at run-time the original guest VM code into an intermediate representation, and then into native code for host machine execution. This feature enables the emulation of different architectures and can be leveraged for tracing and debugging. Our profiler interacts with the translation process to trace the invocation of guest instructions, treating the selected instructions as \emph{events} to trace in our tool. An event of interest is the invocation of a \textit{target function}, representing a fuzzing attack surface, such as an I/O function that handles software input. The target function serves as the entry point for \toolname{}'s fuzzing workflow. During interception, the profiler can also gather input parameters by accessing and reading virtual CPU registers and virtual memory of the guest, facilitating seed collection. The profiler can hook the execution of the target function, either before its execution (\emph{pre-invocation}), by intercepting its first instruction; or, right after its execution (\emph{post-invocation}), by intercepting the instruction at the return address (that was stored at the time of the invocation in the LR register). Event configuration is further discussed in Sec.~\ref{subsec:configuration}.

\bigbreak
\noindent
\textbf{\textit{Injector.}} 
It fuzzes the target function parameters, by accessing and manipulating the state of the guest VM. In particular, fuzz inputs are injected into the virtual CPU registers since arguments are typically passed through general-purpose registers. Depending on the function calling conventions for the CPU architecture, \toolname{} injects fuzz inputs into the proper register. The injector can also modify specific portions of the VM memory, allowing fuzzing of arguments passed through the stack and pointers. For the stack, the injector dereferences memory using the stack pointer register; for pointers, the injector dereferences the register or stack location holding the pointer. \toolname{} requires configuring the function calling conventions only once for the target CPU architecture; then, the same configuration can be reused for fuzzing applications on that architecture. Injections can be performed either pre-invocation or post-invocation, as supported by the \toolname{}, which facilitates fuzzing across various function types. For instance, when dealing with ''\textit{send}'' functions, it is necessary to fuzz the input parameter before the function is invoked. Conversely, in the case of ''\textit{receive}'' functions, it is necessary to fuzz the output parameter after the function has been invoked. This approach helps prevent the fuzzed parameter from being overwritten during the execution of the function, ensuring efficient fuzzing for different function scenarios. After injecting fuzz input, the guest VM resumes the execution.

By intercepting function calls and transparently injecting fuzz input through dynamic binary translation, \toolname{} does not require the introduction of an additional fuzz driver or other modifications to the target. Thus, \toolname{} maintains a completely non-intrusive fuzzing approach, aligning with the first design principle.

\subsection{Feedback}
\label{sec:feedback}
\noindent
\textbf{\textit{Coverage Collector.}}
It supports \toolname{}'s fuzz input generation by transparently gathering \textit{code coverage} from the target during its execution. This process is crucial for effective coverage-guided fuzzing, prioritizing fuzz inputs that explore new code paths while discarding those that do not contribute to exploration. The component analyzes two common types of coverage, \textit{Basic Block Coverage} and \textit{Edge Coverage}, typically used by most gray-box fuzzers. Basic blocks (BBs) represent sequences of instructions bounded by control flow transfers (e.g., jumps, calls, or returns), while edges reveal block-to-block transitions, providing insight into taken paths. \toolname{} performs coverage analysis at the system level, including application code and other components (e.g., the kernel) within the target binary. These additional components can lead to non-deterministic control flow transfers (e.g., due to context switches or timer interrupt handlers), resulting in new spurious code coverage during fuzzing. To address this, the tool conducts a pre-analysis before actual fuzzing, by running the target program with repeated unmodified inputs to blacklist blocks and edges triggered spuriously during this phase. While traditional fuzzers instrument basic blocks at compile-time for coverage tracking, \toolname{} maintains unmodified source and binary code. The coverage collector is integrated into QEMU's translation engine: it uses the program counter of the vCPU and the correspondence BBs and translation blocks (TBs) in the TCG, to gather coverage feedback during execution. This allows \toolname{} to transparently collect both BB and edge coverage without the need for instrumentation support in build toolchains or reliance on hardware-based coverage support. Since fuzzing may involve only a portion of the binary, the bounds to trace are configurable.

\bigbreak
\noindent
\textbf{\textit{Crash Detection.}} \toolname{} enables tracing the occurrence of failure events as additional feedback data, enhancing the input generation of AFL. This feature leverages the same Profiler described in Section~\ref{sec:interception}, which can be optionally configured to intercept crash or error handling functions (e.g., \textit{exit}, \textit{abort}, \textit{assert}). For instance, in MILS-based environments, specific functions responsible for managing virtual board suspension during crashes can be traced.

This novel approach aligns seamlessly with the second and third design principles, emphasizing the flexibility and independence of \toolname{} from specific build configurations and diverse embedded hardware setups.

\subsection{Fuzzing Orchestration}
\label{sec:fuzzing}
\noindent
The orchestrator drives the fuzzing logic in \toolname{}. It serves as a command and control component designed to coordinate the profiler, injector, and coverage collector components, alongside the fuzzer engine (i.e. AFL). \toolname{} implements two different modes of operation: \textit{seed recording} and \textit{intercept-and-fuzz}. 

\bigbreak
\noindent
\textbf{\textit{Seed Recording.}} This mode supports the collection of \textit{fuzz seeds}, that is, initial inputs to be mutated by the fuzzing process. The \textit{orchestrator} uses the \textit{profiler component} to intercept invocations of a configured target function during a regular execution of the target software. The frequency and duration of recording can be configured by the user. The orchestrator gathers a fuzz seed for every interception, which consists of the parameters passed to the intercepted function, and saves them in a queue. Additionally, for each fuzz seed generated, the tool supports the saving of an \textit{initial fuzz state} for snapshot-based fuzzing, which is represented by the VM state at the time of intercepting the function call.

\bigbreak
\noindent
\textbf{\textit{Intercept-and-Fuzz.}} 
This operation mode is designed for event- and time-triggered industrial applications (e.g., MILS-based applications), where software operates based on cyclic internal tasks, which periodically invoke the target function \cite{isovic2000efficient}. 
In this mode, the fuzzing workflow intercepts at run-time the periodic invocations of the target function, and replaces the original inputs with fuzz inputs. The \textit{orchestrator} selects an input to be fuzzed from the pre-recorded \textit{fuzz seed}. The \textit{fuzzer engine} applies mutations to the seed (e.g., bit flips, byte flips, arithmetic operations, and havoc as in the AFL fuzzer), generating a set of \textit{fuzz inputs}, which are enqueued. As the target software runs on the guest VM, the \textit{profiler} intercepts regular invocations of the target function. For each invocation intercepted during the fuzzing campaign, the orchestrator selects a fuzz input from the queue, and injects it using the \textit{injector}. Throughout the campaign, the \textit{orchestrator} utilizes the \textit{coverage collector} to inform the fuzzer engine about new execution paths. Fuzz inputs exploring new paths are marked as interesting and included in the queue for mutation. Additionally, during the campaign, the orchestrator also uses the profiler to identify anomalous behaviors, such as VM crashes or hangs by tracking abort, exit, or assert functions. The detected states serve as test outcomes for further analysis. The user terminates the fuzzing campaign.

\subsection{Configuration}
\label{subsec:configuration}
Since MILS systems can support many diverse and complex system configurations, \toolname{} is highly configurable to adapt and perform fuzzing on different target operations. Then, it requires specific configuration parameters for effective usage. The target function for fuzzing can be kernel system calls, or library function calls in a user-space application. For instance, when fuzzing MILS-based applications (e.g., an application gateway), \toolname{} can focus on SIPC kernel primitives for inter-partition communication. Users need to identify the target function and configure \toolname{} with the associated symbol. We assume that information about the target function is available to the user through product documentation and specifications (e.g., documentation from the vendor of the MILS kernel about kernel primitives), and technical standards (e.g., the POSIX standard used by UNIX, Linux, and other OSes). \toolname{} then parses the target binary (e.g., in ELF format) to automatically identify the memory address where target symbols will be loaded, using the \textit{GNU nm} utility. The locations of the symbols become the entry points for fuzzing. If symbol information is stripped from the binary, the user requires additional solutions to identify the addresses, which have been developed in other studies and are outside the scope of this work. The user can adopt reverse engineering tools like Fuzzable \cite{fuzzable} to automate the discovery of the target function address through static analysis on binaries.

The user can specify which input parameters of the function call are going to be fuzzed. Fuzzing often targets complex input data (e.g., large data structures), which are typically passed to functions as pointer parameters. In our current design, the tool dereferences the indicated input parameter as a pointer, and corrupts the pointed memory area. The size of the memory area to be corrupted can be specified in two ways: it can be statically configured by the user (e.g., when fuzzing a \texttt{struct} parameter used for passing complex data structure in C), or it can be derived from another input parameter provided to the function (e.g., C functions often pass byte arrays along with an integer parameter indicating the array size), which the user can also configure.

\toolname{} is adaptable to various CPU architectures, by allowing the user to configure the calling conventions and register usage specific to the target architecture. For example, the PowerPC architecture uses the general-purpose registers GPR3-GPR10 to pass the first eight parameters, while any remaining parameters are passed on the stack \cite{powerPCconvention}. Similarly, ARM and MIPS pass the first four parameters in registers R0-R3 and A0-A0, respectively. In addition, the tool can retrieve the return address of a function call at the time of the invocation, such as, by reading the Link Register (LR) register in the ARM and PowerPC architectures, or from the Return Address Register (RA) in MIPS architecture. This architectural awareness ensures that \toolname{} accurately intercepts functions and parameters and injects the fuzz input. 

In addition, during the configuration phase, utilizing the Code Coverage Collector component allows running the target system in idle conditions, facilitating the automatic gathering of basic blocks to blacklist for the coverage feedback, as discussed in Section \ref{sec:feedback}.

The configuration of the information described in this section represents the only manual effort required to use \toolname{}. The calling conventions need to be configured just once for each architecture, and \toolname{} comes with pre-defined profiles for popular architectures including PowerPC, ARM, and MIPS. The configuration of the target function is performed only once for the target binary, and can potentially be reused across different applications that use the same kernel and functions (e.g., the SIPC primitives of a MILS kernel product). While setup time varies depending on the target system and user expertise, configuring a new target typically requires between 30 minutes and a few hours. In our experience, most of this time is spent reviewing target specifications and header files, and is significantly less than the effort needed to develop a dedicated fuzz driver to the target application, which is a key limitation towards the adoption of fuzzing \cite{babic2019fudge,ispoglou2020fuzzgen,jeong2023utopia}.

\section{Implementation}
\label{sec:implementation}
\toolname{} is built upon QEMU emulator version 7.0. We chose QEMU due to its broad support for several CPU architectures, including x86, ARM, MIPS, and PowerPC \cite{qemu_architecture_support}. Additionally, it offers support for various physical boards and peripherals, and its inherent extensibility further adds to its potential. In fact, we have added customization gathered from Adacore \cite{adacore_qemu} (a partner of VxWorks) to support the MPC8548E development board, which is the System-on-Chip on which VxWorks MILS executes. Our implementation of \toolname{} is integrated into the codebase of the emulation engine. The fuzzing engine within \toolname{} is developed using \emph{libAFL} \cite{fioraldi2022libafl}, a modular and configurable library that incorporates AFL algorithms. The details of each component are briefly described below.

\bigbreak 
\noindent
\textbf{\textit{Profiler.}}
It is implemented as a TCG plugin \cite{TCG_plugin}, which subscribes to events generated by the QEMU translation process. In particular, the profiler registers a callback to hook the translation of each new basic block. If the translated block includes the instruction marked as the fuzzing entry point (from the configuration), a second callback is registered to hook the execution of the entry point. Upon the entry point execution, the plugin pauses the VM, initiating VM state profiling to collect target function parameters. While TCG plugins in QEMU can passively monitor the system by logging aspects such as executed instructions or memory accesses \cite{TCG_plugin}, they currently lack the ability to read the guest state during execution. To address this limitation, we extended the QEMU API with two custom functions: '\texttt{get\_cpu\_register()}' and '\texttt{vcpu\_read\_phys\_mem()}', providing access to guest vCPU registers and virtual memory, respectively. These functions interact with internal QEMU data structures, and our profiler can store this data in host-side files, simplifying seed collection.

\bigbreak
\noindent
\textbf{\textit{Injector.}}
It is implemented as an additional configurable callback within the same TCG plugin, triggered sequentially after the profiler, during fuzzing. To facilitate the fuzzing of target function parameters, we expanded the QEMU API with two additional custom functions to override guest vCPU registers and memory.

\bigbreak
\noindent
\textbf{\textit{Coverage Collector.}} 
It is integrated into QEMU's translation engine to collect \textit{Block} and \textit{Edge coverage} from embedded binaries executed on \toolname{}. This implementation is based on AFL's QEMU mode, a QEMU patch originally designed for capturing an edge-trace in user space Linux binaries. \toolname{} extends this mode to support binaries running in QEMU's full-system emulation, enabling the tracing of embedded applications, kernels, and firmware.

\bigbreak
\noindent
\textbf{\textit{Fuzzing Orchestrator.}}
It runs as a separate thread in the QEMU process. This thread controls and synchronizes all other architectural components, as well as the fuzzing engine (which runs in a second QEMU thread). Additionally, it handles the parsing of user parameters from the command-line interface (CLI) and configures \toolname{} accordingly.

\section{Evaluation on Industrial Embedded Systems}
\label{sec:evaluation}

In this section, we evaluate \toolname{}'s effectiveness in enabling modern gray-box fuzzing for binary-only targets that rely on proprietary toolchains and platforms, as found in industrial systems. Our evaluation needs to focus on the architectural advantages of \toolname{}, while avoiding biases due to different fuzz test generation techniques. This is the reason why we introduce our own baseline approach, which uses the same LibAFL backend for input generation as \toolname{}. The key difference between the approaches lies in the architecture, since the baseline must rely on a traditional setup with a custom fuzz driver and liveness checks, without leveraging code instrumentation.

Our evaluation is structured around two primary aspects: fuzzing depth and performance. Fuzzing depth is assessed through bug discovery and code coverage. Bug discovery measures the ability of \toolname{} to uncover potential vulnerabilities, while code coverage measures the extent to which the tool explores execution paths in the target, indicating the comprehensiveness of its analysis. Performance is evaluated in terms of throughput, defined as the number of fuzz inputs processed per second. Since fuzzing requires the generation and execution of a large volume of inputs, high throughput is essential to maintain efficiency and scalability. 

\subsection{Evaluation Targets}
\label{sec:experimental_methodology}
In our experiments, we consider the Wind River VxWorks MILS industrial platform. These experiments consider MILS-based applications as targets of evaluation. The MILS-based targets selected for our evaluation are inspired by a real-world defense domain gateway application, which involves client-server communication where \textit{parsing libraries} are used to validate and filter exchanged payloads. These systems are of critical importance in security-sensitive environments, but testing them poses significant challenges (see Section \ref{sec:motivating_example}), and existing fuzzers cannot be applied effectively. To replicate the functionalities of a defense-domain gateway we used the VxWorks MILS hypervisor alongside open-source (vulnerable) software components, since we are unable to disclose information about the actual proprietary application. MILS guard applications perform input validation and parsing of data inputs, in order to inspect the data before they are forwarded to another network. 

Therefore, we consider libraries that parse structured data formats as fuzzing targets. We selected three POSIX-compliant open-source parsing libraries, with minimal external dependencies, in order to facilitate the porting of these libraries to VxWorks MILS. We consider JsonParser \cite{JsonParser}, responsible for parsing serial data as JSON strings; SendMail \cite{SendMail}, a preprocessor for email addresses aimed at normalization; TinyExpr \cite{TinyExpr}, utilized for parsing and evaluating mathematical expressions. These targets perform parsing and validation of diverse structured data formats, similarly to real-world security guard applications \cite{rushby1981design}. Vulnerabilities in such libraries could compromise the security guarantees provided by the MILS architecture. We compiled binary executable using the original toolchain from the MILS product, without any change to the build process. We performed the experiments without assuming any access to the source code and to the build toolchain. The system was run on a QEMU virtual machine emulating the \textit{WindRiver SBC8548E} development board, which uses a PowerPC MPC8548E CPU with 1 GB RAM.

Table \ref{tab:mils_target} lists the selected targets, their type, and the CPU architectures that were emulated during the experiments. We conducted experiments on a host system with an \textit{AMD Ryzen 5 3600 6-Core 3.98 GHz} processor and 12GB RAM. 

\begin{table}[h]
\centering                    
\scriptsize
\caption{MILS-based Applications Targeted in the Evaluation.}
\label{tab:mils_target}
\begin{tabularx}{\columnwidth}{lll}
  \toprule
  \textbf{Target} & \textbf{Description} & \textbf{Architecture} \\
  \midrule
  json \cite{JsonParser} & Parses serial data as JSON strings & PowerPC \\
  sendmail \cite{SendMail} & Pre-processes email addresses & PowerPC \\
  tinyexpr \cite{TinyExpr} & Parses mathematical expressions & PowerPC \\
  \bottomrule
\end{tabularx}
\end{table}

\subsection{Experimental Setup}
To evaluate and compare \toolname{} coverage-driven approach against traditional approaches, we configured two distinct testing setup.

\begin{figure*}[h]
  \centering
  \includegraphics[width=0.6\linewidth]{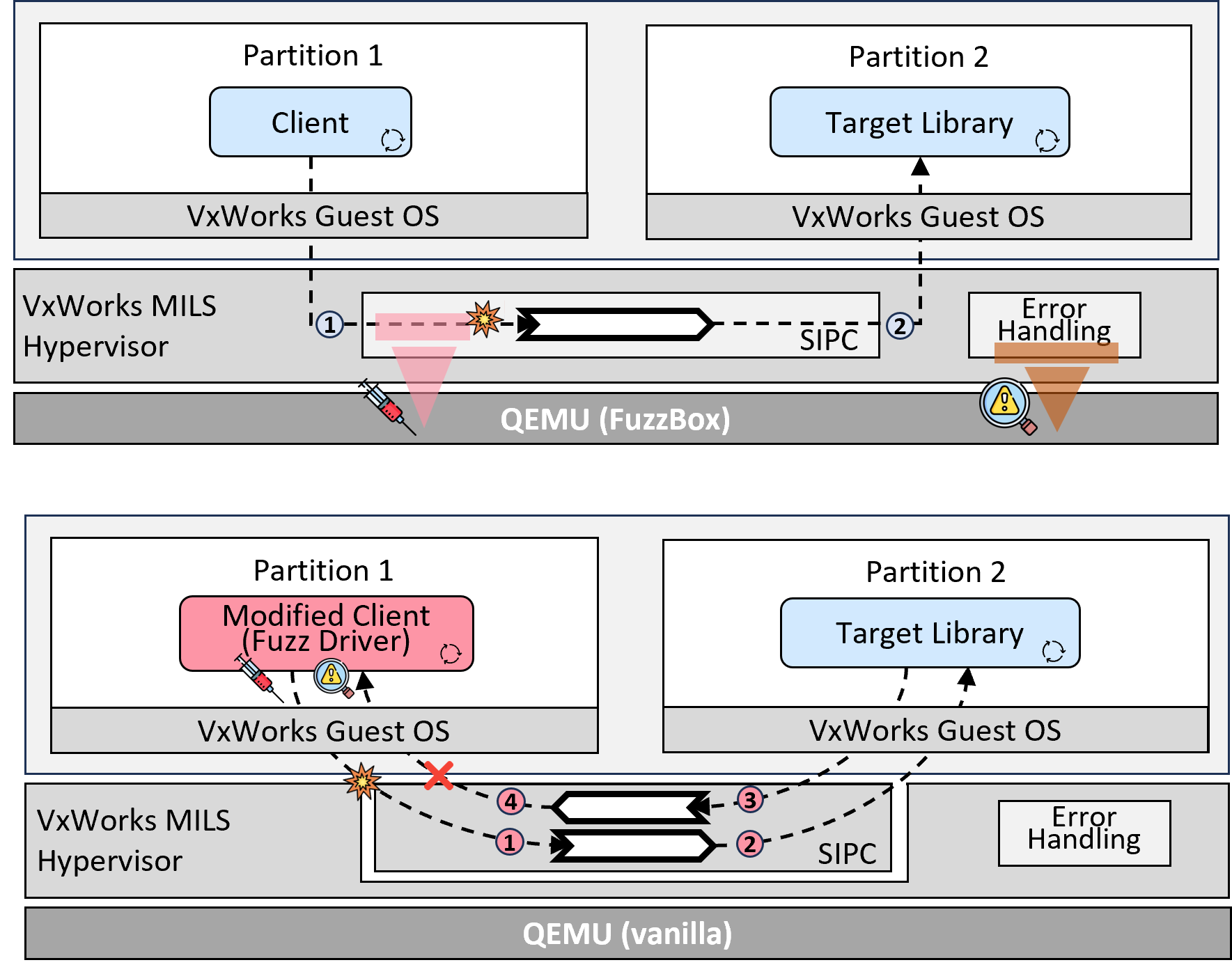}
  \caption{\toolname{} (top) and \textit{baseline} (bottom) setup to test MILS systems.}
  \label{fig:experimental_setup}
\end{figure*}

\bigbreak
\noindent
\textbf{\toolname{} \textit{Testing Setup}.} 
We configured VxWorks MILS to implement the MILS gateway use case in a simplified setup. Figure \ref{fig:experimental_setup} (top) shows the \toolname{} testing setup, involving two VBs running the VxWorks Guest OS. These VBs communicate through SIPC channels through the VxWorks MILS hypervisor. 
The client uses the \texttt{sendMessageSIPC()} kernel primitive to iteratively send messages to the second VB (step \circled{1}). 

The server uses the \texttt{receiveMessageSIPC()} routine to receive messages and invokes the target library to process them (step \circled{2}). In addition, in this setup, the MILS hypervisor utilizes \texttt{schedSuspendVb()} primitive to manage the suspension of VBs during crashes, ensuring fail-safe behavior. We compiled all of the described components and workflows in a binary package that we executed on top of a \toolname{}-enabled VM. In this testing configuration, \toolname{} operates in \textit{intercept-and-fuzz} mode with edge coverage feedback enabled. Configuring \toolname{} for a specific target to fuzz is a straightforward process, ensuring a user-friendly experience with minimal manual effort. The user identifies the target function (\texttt{sendMessageSIPC()}), specifies the parameters to fuzz and their position (3rd and 4th parameters as message and length), selects the CPU architecture (PowerPC), and chooses the crash detection function (\texttt{schedSuspendVb()}).

\bigbreak
\noindent
\textbf{\textit{Baseline Testing Setup}.} 
The Baseline setup adopts a traditional configuration (the basic fuzzing workflow outlined in Section \ref{sec:motivating_example}), with a modified client that acts as a custom fuzz driver and failure detector through liveness checks. 
In this setup, we consider the popular libAFL library \cite{fioraldi2022libafl} as reference fuzzer. We emphasize that both the \toolname{} and the Baseline setups use the same LibAFL-based fuzzing engine, with identical input mutation logic and seeds. This ensures that any observed differences in fuzzing depth or throughput can be attributed to architectural differences, particularly the enhanced introspection and feedback loop enabled by \toolname{}, rather than to different fuzz input generation strategies.

We remark that we could not consider other state-of-the-art fuzzers for our evaluation on MILS-based applications, as they lack support for niche operating systems and architectures used in industrial systems (see the challenges identified in Section~\ref{related_work}).

Therefore, we had to develop a custom fuzz driver, using the libAFL library to orchestrate fuzzing (e.g., generating mutations, enqueueing fuzz inputs, etc.). Figure \ref{fig:experimental_setup} (right) shows the Baseline configuration. Again, two VBs communicate through an SIPC channel. In this setup, a black-box fuzz driver is integrated into the application, taking the role of the client. It sends fuzz inputs to the server in the second VB, which is the same as the previous setup (steps \circled{1} and \circled{2}). Since there is no direct way for the client VB to check the status of the server VB, the setup also includes an additional SIPC channel (not needed by the MILS-based gateway application) for response messages from the server to the client. The fuzz driver checks the response queue from the server, by setting a timeout to detect failures (steps \circled{3} and \circled{4}). The resulting binary is executed on a vanilla QEMU, with all \toolname{} components disabled except for the coverage collector, used for experimental evaluation purposes. \textit{Baseline} is limited to collect coverage statistics, without using feedback to guide the fuzzing process. Without our tool, this setup is only feasible when it is possible to recompile the target application, with an additional effort to adapt the fuzz driver.

\subsection{Fuzzing depth}
\toolname{} enables coverage-guided fuzzing for binary-only industrial targets, where traditional coverage measurement is challenging due to limited introspection capabilities. To evaluate the fuzzing depth achieved by our proposed technique, we compared its code coverage and bug finding capabilities against traditional black-box fuzzing \cite{klees2018evaluating}. 

In both configurations, the \toolname{} architecture was used to collect code coverage metrics. However, in the black-box fuzzing setting, coverage data was gathered solely for measurement purposes and did not influence the fuzzing input generation. In contrast, \toolname{} setup used the coverage feedback to guide the input generation process.

To assess the bug finding effectiveness of \toolname{} against the \textit{Baseline}, we built two buggy versions of each of the three POSIX-compliant libraries introduced in Section \ref{sec:experimental_methodology}, utilizing two different injection strategies, resulting in six evaluation targets. The ``\textit{easy}'' buggy versions of the libraries include a bug injected into frequently-exercised and easy-to-reach path of the application, discoverable with straightforward fuzz inputs and minimal mutations. Conversely, the ``\textit{hard}'' buggy versions include a bug injected deeper within the application logic, requiring a more intricate fuzz mutations for discovery. This approach is inspired by the well-known LAVA approach \cite{dolan2016lava}, which injects bugs into carefully-selected paths of a program to challenge bug finding tools. We ensured that each vulnerability has an independent input trigger condition. The injected bugs cause a memory violation, which is representative of real-world memory safety vulnerabilities in embedded systems \cite{anatomy2018, Characterizing2013, geng2020empirical}. Specifically, the bug corrupts a register of the TSEC Ethernet controller in a way that triggers a read from a privileged memory region. This region lies outside the allowed address space of the application code running on the virtual board, and any access to it results in a crash of the VB itself. This fault reflects a common class of vulnerabilities in embedded systems, where misconfigured or attacker-controlled peripheral accesses can violate memory protection boundaries enforced by the RTOS and hardware MMU. These bugs mirror known vulnerability patterns in MILS-based systems, such as CVE-2019-12255 and CVE-2019-12264, where malformed inputs to the network stack of VxWorks OS led to memory accesses outside the permissible range, involving control and status registers mapped to privileged memory.

\begin{figure*}[h]
  \includegraphics[width=\textwidth]{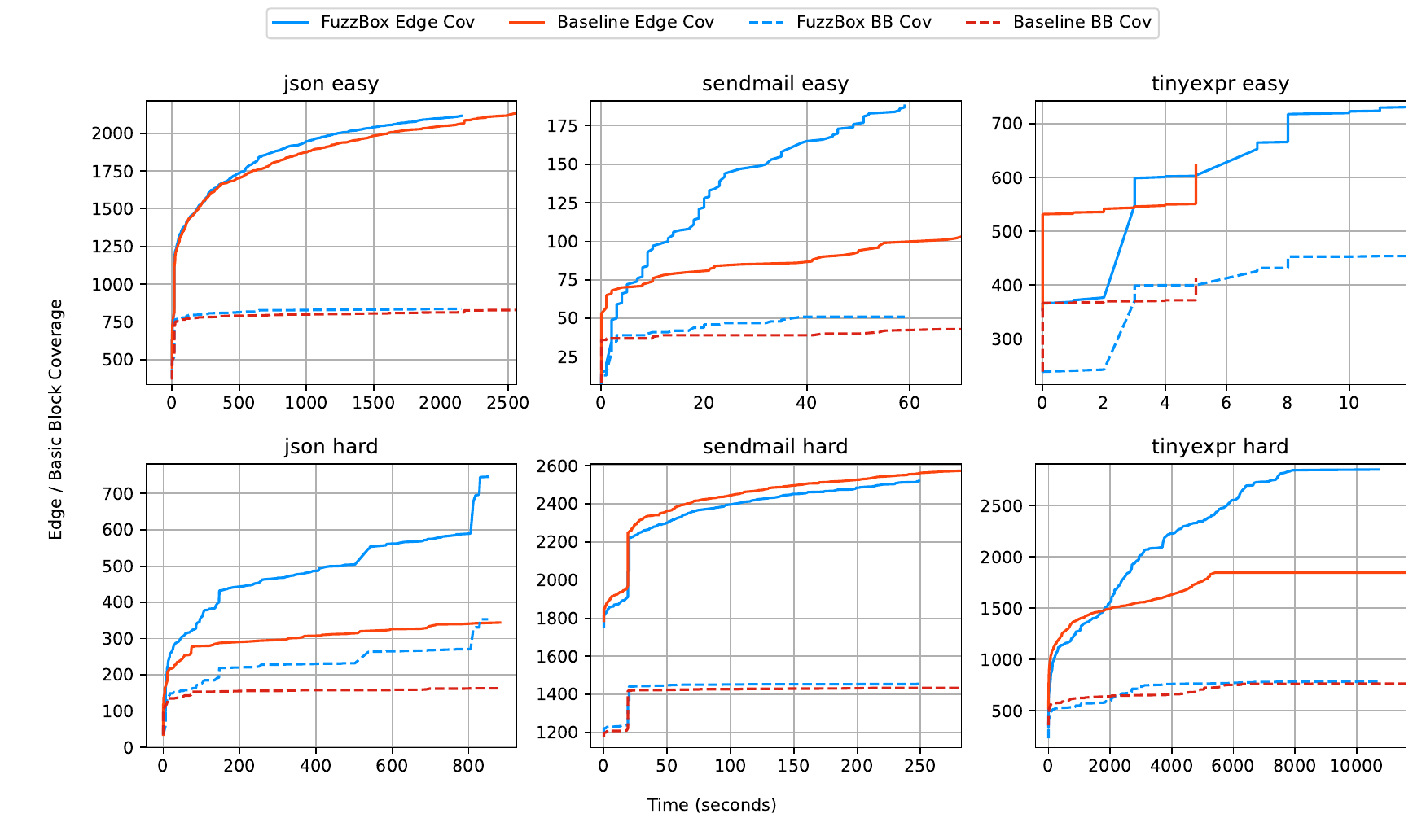}
  \caption{Coverage growth curves over time for MILS-based application targets.}
  \label{fig:coverage}
\end{figure*}

\bigbreak
\noindent
\textbf{\textit{Coverage Analysis.}}
In the first experiment, we evaluated the capability of \toolname{} to achieve high code coverage compared to the \textit{Baseline} approach. Since the two approaches may trigger a bug in the target within different time windows, with one potentially requiring less time than the other, we standardized the comparison by fixing the time window to the shorter duration required to trigger the bug between the two approaches. Then we measured edge and basic block coverages for both approaches in the selected time window. Table \ref{tab:coverage_statistics} shows the total number of edges and blocks discovered by the two techniques in the reference time windows, across the selected targets. Furthermore, both coverage growth curves over time are plotted in Figure \ref{fig:coverage}. 

\begin{table}[h!]
\caption{Edge (and BB) coverage achieved by \toolname{} and \textit{Baseline} in a fixed time window. }
\label{tab:coverage_statistics}
\centering
\scriptsize
\begin{tabular}{p{0.15\linewidth}  p{0.15\linewidth} p{0.17\linewidth}  p{0.17\linewidth}  p{0.15\linewidth}}
  \toprule
  \textbf{Target} & \textbf{T.W. (s)} &\textbf{FuzzBox}  & \textbf{Baseline} & \textbf{Impr \% } \\
  \midrule
  \textit{json\_easy}       & 2197.53   & 2117 (837) & 1913 (801) & +10.66\% \\
  \textit{json\_hard}       & 895.09    & 746 (353) & 327 (159) & +128.13\% \\
  \textit{sendmail\_easy}   & 85.22     & 188 (51) & 84 (39) & +123.80\% \\
  \textit{sendmail\_hard}   & 282.68    & 2520 (1455) & 2559 (1434) & -1.52\% \\
  \textit{tinyexpr\_easy}   & 6.12      &653 (426) & 622 (413) & +4.98\% \\
  \textit{tinyexpr\_hard}   & 11913.52  &2851 (782) & 1600 (654) & +78.18\% \\
  \bottomrule
\end{tabular}
\end{table}

Each test concludes when a crash is identified in the target or when coverage saturation is observed. Indeed, using coverage as feedback to generate fuzz inputs improves the discovery of new paths in the program. Examining the final edge coverage achieved by \toolname{} and the \textit{Baseline}, our solution demonstrates changes of +10.66\%, +128.13\%, +123.80\%, -1.52\%, +4.98\%, and +78.18\% across the six targets, with an average improvement of 57.37\%. Furthermore, analyzing the coverage curves over time, in most cases, \toolname{} achieves higher edge and basic block coverage more rapidly than the \textit{Baseline}.

\verbdef\seedjsoneasy|AA|
\verbdef\seedjsonhard|{bbbbbbb}|
\verbdef\seedmaileasy|AAAAA|
\verbdef\seedmailhard|ABCDEFG|
\verbdef\seedtinyeasy|help(|
\verbdef\seedtinyhard|1+2|

\verbdef\stringajsoneasy|{}|
\verbdef\stringajsonhardA|{"^?\bpp|
\verbdef\stringajsonhardB|{"bi\b|

\verbdef\stringasmtpeasyA|(B((^],\Û^i(BÛ^i(B^c×œ×Bd|
\verbdef\stringasmtphardA|X@A|
\verbdef\stringasmtphardB|X@g>A|

\verbdef\stringatinyeasyA|help|
\verbdef\stringatinyhardB|e+e...(etc)|

\verbdef\num|#|

\begin{table*}[t]
  \centering
  \scriptsize
  \caption{Discovered bugs and TTC in \toolname{} and \textit{Baseline} for MILS-based applications.}
  \label{tab:bug_discovery}
  
  \begin{tabular}{p{0.14\linewidth} p{0.10\linewidth} p{0.25\linewidth} p{0.11\linewidth} p{0.11\linewidth} p{0.10\linewidth}}
    \toprule
    \textbf{Target} & \textbf{Seed} & \textbf{Fuzz Input} & \textbf{TTC (s)} & \textbf{Iter.} & \textbf{Approach} \\
    \midrule

    \multirow{2}{*}{\textit{json\_easy}} & \multirow{2}{*}{\seedjsoneasy} & \stringajsoneasy & \cellcolor{green!30} $2197.53$  & \cellcolor{green!30} 95481 & \cellcolor{green!30} \toolname{}\\
               &                & \stringajsoneasy & \cellcolor{red!30} $4987.66$ & \cellcolor{red!30} 397629 & \cellcolor{red!30} \textit{Baseline}\\

    \arrayrulecolor{customgrey}\hline\arrayrulecolor{black}
    
    \multirow{2}{*}{\textit{json\_hard}}  & \multirow{2}{*}{\seedjsonhard} & \stringajsonhardA & \cellcolor{green!30} $895.09$  & \cellcolor{green!30} 53100 & \cellcolor{green!30} \toolname{}\\
                &                & \stringajsonhardB & \cellcolor{red!30} $921.42$ & \cellcolor{red!30} 71173 &\cellcolor{red!30} \textit{Baseline}\\

    \arrayrulecolor{customgrey}\hline\arrayrulecolor{black}
    
    \multirow{2}{*}{\textit{sendmail\_easy}}  & \multirow{2}{*}{\seedmailhard}  & \stringasmtpeasyA & \cellcolor{green!30} $85.22$ & \cellcolor{green!30} 2817  & \cellcolor{green!30} \toolname{}\\
                &                & -                & \cellcolor{red!30} $>$ 1h & \cellcolor{red!30} 200000+ &\cellcolor{red!30} \textit{Baseline}\\

    \arrayrulecolor{customgrey}\hline\arrayrulecolor{black}
    
    \multirow{2}{*}{\textit{sendmail\_hard}} &  \multirow{2}{*}{\seedmaileasy}  & \stringasmtphardA & \cellcolor{green!30} $282.68$ & \cellcolor{green!30} 16960 & \cellcolor{green!30} \toolname{}\\
               &                 & \stringasmtphardB & \cellcolor{red!30} $461.45$ & \cellcolor{red!30} 54963 & \cellcolor{red!30} \textit{Baseline}\\
    
    \arrayrulecolor{customgrey}\hline\arrayrulecolor{black}
    
    \multirow{2}{*}{\textit{tinyexpr\_easy}} & \multirow{2}{*}{\seedtinyeasy} & \stringatinyeasyA & \cellcolor{red!30} $1001.51$   & \cellcolor{red!30} 5028 & \cellcolor{red!30} \toolname{}\\
                   &                & \stringatinyeasyA & \cellcolor{green!30} $6.12$ & \cellcolor{green!30} 339 & \cellcolor{green!30} \textit{Baseline}\\

    \arrayrulecolor{customgrey}\hline\arrayrulecolor{black}
    
    \multirow{2}{*}{\textit{tinyexpr\_hard}} & \multirow{2}{*}{\seedtinyhard}  & \stringatinyhardB & \cellcolor{green!30} $11913.52$ & \cellcolor{green!30} 209312 & \cellcolor{green!30} \toolname{}\\
                    &              & -                 & \cellcolor{red!30}  $>$ 3h   & \cellcolor{red!30} 300000+ & \cellcolor{red!30}  \textit{Baseline}\\
    \bottomrule
  \end{tabular}
\end{table*}

\bigbreak
\noindent
\textbf{\textit{Bug Analysis.}}
In a second experiment, we evaluated the bug discovery capabilities of \toolname{}. Table \ref{tab:bug_discovery} presents the results for the six targets. For each case, it reports the time to crash (TTC) in seconds and the number of iterations required (i.e., unique mutations) to cause a crash. Depending on the target, different initial seeds were used, as documented in the table. The determinism of AFL ensures the consistent generation of the same input sequences when using the same seed in multiple test repetitions. This makes the tests reproducible and eliminates the need for statistical measures that assess variability, such as reporting multiple repetitions or calculating average times. When applied to the ''\textit{hard}'' targets, \toolname{} consistently outperforms the \textit{baseline} approach. In two cases, the \textit{baseline} fails to detect the bug within a 1-hour time window. This highlights the benefit of leveraging code coverage, obtained by our tool through the QEMU translation process. In a single ''\textit{easy}'' case, the black box \textit{baseline} approach is quicker than \toolname{} since simple mutations suffice to trigger the bug. As \toolname{} adopts a gray-box approach, the input generation algorithm focuses on more complex combinations of multiple mutations, to better explore the target code paths. This approach is more effective for hard-to-find bugs and for achieving higher coverage in the long term, even if it is overblown for simpler bugs.

\bigbreak
\subsection{Performance}
We assessed the performance of \toolname{} and compared it with the \textit{Baseline}, in terms of throughput of fuzzing (i.e., executions of fuzz inputs per second) when applied against MILS-based targets. We separately analyze the three libraries, considering potential variations among the libraries with respect to the time needed to process the inputs. In each experiment, we first submit 100 fuzz inputs to warm up the target, in order to mitigate initial transient variations in the measurements, due to the initialization of the fuzzer. Subsequently, we measured throughput with an additional 1000 fuzz inputs. To ensure statistical significance, we performed three repetitions for each experiment. Table \ref{tab:throughput} provides the average throughput, along with the standard deviation, for both approaches and each target. The results show that, for all three targets, the throughput achieved by \toolname{} is approximately twice that of the \textit{Baseline}. This underscores that the overhead introduced by \toolname{} for profiling, injecting, and instrumenting coverage within the translation process is minimal compared to the overhead associated with a naive fuzzing workflow. Given potential transient slowdowns in the target system (e.g., with large fuzz inputs), we properly set timeouts on the client side to await server liveness messages, long enough to prevent false positives in failure detection. Nevertheless, this configuration significantly affects the overhead on the Baseline fuzzing throughput.

\begin{table}[h]
\centering
\scriptsize
\caption{Throughput of FuzzBox compared to the Baseline.}
\label{tab:throughput}
\begin{tabularx}{\columnwidth}{>{\raggedright\arraybackslash}X 
                                 >{\raggedright\arraybackslash}X 
                                 >{\raggedright\arraybackslash}X}
  \toprule
  \textbf{Target} & \textbf{Baseline Throughput (\#input/sec)} & \textbf{\toolname{} Throughput (\#input/sec)} \\
  \midrule
  \textit{json}     & $23.05 \pm 0.19$ & $47.92 \pm 1.72$ \\
  \textit{sendmail} & $24.65 \pm 0.93$ & $53.09 \pm 3.53$ \\
  \textit{tinyexpr} & $20.90 \pm 1.55$ & $40.68 \pm 0.78$ \\
  \bottomrule
\end{tabularx}
\end{table}

\bigbreak
\Definition{
This evaluation highlights \toolname{} as a groundbreaking solution for fuzzing industrial embedded systems like MILS-based applications. Unlike the baseline, which relies on custom fuzz drivers and recompilation with limited introspection, \toolname{} achieves superior fuzzing depth, discovering hard-to-reach bugs with a 57.37\% average coverage improvement. Additionally, it delivers nearly twice the throughput, minimizing the overhead caused by liveness-based crash detection. These results position \toolname{} as a high-performance, effective tool for addressing the unique challenges of fuzzing binary-only industrial targets.
}

\section{Portability Evaluation}
\label{sec:iot}

In this section, we evaluate the capability of \toolname{} to fuzz targets beyond MILS-based systems. To assess its portability and broader applicability, we applied \toolname{} to a diverse set of Linux-based IoT firmware from commercial embedded devices. Additionally, we compared the improvements achieved by \toolname{} to a \textit{Baseline} approach based on black-box network fuzzing. While some of the existing state-of-the-art fuzzers are applicable to IoT systems, the goal of this evaluation is not to compare \toolname{} against those tools directly. These fuzzers adopt their own fuzz input generation strategies, which would make it difficult to attribute any observed differences in performance solely to architectural factors. To ensure a fair and controlled comparison, we use a Baseline setup that shares the same LibAFL backend as \toolname{}. Since our goal is to validate the general applicability of FuzzBox across different embedded system types, CPU architectures, and operating systems, the Baseline setup assures us that experiments use identical input mutation and scheduling strategies.

\subsection{Evaluation Targets}
These experiments consider Linux-based IoT firmware as targets of evaluation. Unlike MILS systems, IoT firmware runs atop general-purpose Linux operating systems and has been extensively studied as fuzzing targets in the literature \cite{kim2020firmae, kim2021firm, zheng2019firm}. For this evaluation, we selected real-word firmware of three popular IoT devices. We included dual-band gigabit WiFi routers and IP cameras, running on embedded architectures based on ARM and MIPS CPUs. 

Table \ref{tab:firm_target} lists the selected targets, their type, and the CPU architectures that were emulated during the experiments. 
We executed the firmware on QEMU VMs with 1 GB RAM, running on a host system with an \textit{AMD Ryzen 5 3600 6-Core 3.98 GHz} processor and 12GB RAM.

\begin{table}[h!]
\centering                    
\scriptsize
\caption{Linux-based Firmware for portability evaluation.}
\label{tab:firm_target}
\begin{tabularx}{\columnwidth}{lll}
  \toprule
  \textbf{Target} & \textbf{Description} & \textbf{Architecture} \\
  \midrule
  TENDA AC15 \cite{tenda_ac15} & WiFi router firmware & ARM (little endian) \\
  TEW-651BR \cite{tew_firmware}  & WiFi router firmware & MIPS (big endian) \\
  DCS-932L \cite{DSC_firmware}   & IP camera firmware & MIPS (little endian) \\
  %ASUS RT-N53 \cite{x} & WiFi router firmware & MIPS \\
  \bottomrule
\end{tabularx}
\end{table}

\subsection{Experimental Setup}
To evaluate and compare \toolname{} coverage-driven approach against traditional network fuzzing approach, we configured two distinct testing setup.

\begin{figure*}[h]
\centering
  \includegraphics[width=1\linewidth]{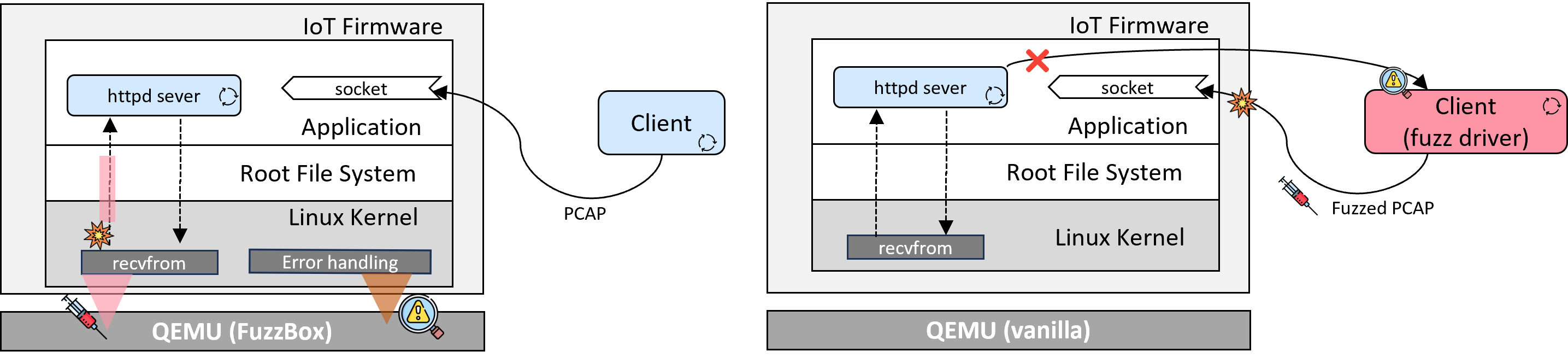}
  \caption{\textit{Firmware FuzzBox} (left) and \textit{Firmware baseline} (right) testing setup. }
  \label{fig:experimental_setup_firm}
\end{figure*}

\bigbreak
\noindent
\textbf{\toolname{} \textit{Testing Setup}.} 
For the evaluation of \toolname{} on the IoT firmware target, we setup a Linux-based environment. As shown in Figure \ref{fig:experimental_setup_firm} (left), the Linux-based firmware runs a \textit{httpd} server, actively listening for incoming HTTP requests related to network configuration, security settings, and communication protocols. Upon receiving an HTTP request through the network interface, the firmware utilizes the \texttt{recvfrom()} Linux system call to process incoming packets from a socket. In contrast to MILS-based applications, where invocation cycles are periodically triggered, the target firmware requires external stimulation. To achieve this, we included an external client to send a pre-recorded seed in PCAP format as a repeated stream of packets to the server, triggering the functions we aim to intercept and fuzz (i.e., \texttt{recvfrom()}). This eliminates the need for a protocol-specific fuzz client, as required in other fuzzing approaches \cite{pham2020aflnet}. 

In order to attack the firmware, we fuzz messages over the socket before they are received by the target application.  
To enable this, \toolname{} intercepts the receiving system call, both before and after its execution (pre- and post-invocation). During pre-invocation, \toolname{} collects information about the memory location designated for saving the received messages, and saves the return address of the call. At post-invocation, when the execution reaches the return address, \toolname{} performs the actual fuzzing operation on the messages. 

Thus, when fuzzing received messages, \toolname{} manipulates the buffer of the \texttt{recvfrom()} syscall. Since blindly applying mutations across the entire HTTP payload would be inefficient, fuzzing tools for HTTP, such as the Burp Suite \cite{burpsuite} and OWASP ZAP \cite{owasp_zap}, are configured to selectively mutate specific fields in HTTP requests, either in the header or body (e.g., parameters in the HTTP query string). To address this, \toolname{} supports additional configuration parameters, including an offset and the length to fuzz for selective fuzzing of specific bytes within the syscall input/output buffer. This allows users to target specific parts of intercepted HTTP requests for more efficient fuzzing.

Additionally, for Linux-based targets, crash detection in \toolname{} is achieved by intercepting Linux core dumps, which are typically generated when a program encounters a segmentation fault or another critical error for the running target application. 

In summary, the \toolname{} configuration parameters for this setup include setting \texttt{recvfrom} as the target syscall, specifying its parameter ''\textit{buf}'' (at the second position) for fuzzing, and defining the offset and the length within which to fuzz the buffer. The \texttt{do\_coredump()} signal handler is configured to intercept and detect firmware crashes. Finally, the target CPU architecture is selected, as the parameter to fuzz is passed in different CPU registers depending on the architecture (e.g., the 2nd parameter of a syscall is passed via the GPR4 register in PowerPC CPUs).

\bigbreak
\noindent
\textbf{\textit{Baseline Testing Setup.}}
To provide a basis for comparison with \toolname{}, we configured a baseline setup. This configuration, shown in Figure \ref{fig:experimental_setup_firm} (bottom), employs a simple black-box fuzzing approach for IoT firmware. In this setup, a fuzz driver is integrated into the external client, which uses libAFL to generate and schedule fuzzing inputs by mutating the pre-recorded PCAP seed. Users can specify which fields to mutate by applying a mask to selected bytes within the PCAP file. This mechanism enables precise, selective mutation, achieving a granularity level comparable to that of the \toolname{} testing setup. The fuzz driver sends these mutated inputs to the target server, which processes them in the same manner as in the \toolname{} setup. The fuzz driver then waits for a response from the server, using a timeout mechanism to detect crashes by flagging unresponsive states.

The firmware image, comprising a Linux kernel, root filesystem, and the target application, is executed on a vanilla QEMU instance. Similar to the MILS-based setup, coverage collection of \toolname{} is enabled for the evaluation and not utilized as feedback mechanism.

This baseline configuration reflects traditional black-box network fuzzing techniques commonly employed for IoT firmware \cite{zheng2019firm, chen2018iotfuzzer, feng2021snipuzz}. In these techniques, the fuzzer operates remotely, communicates with the target through network interfaces, and relies solely on analyzing response messages, without utilizing others feedback mechanisms. However, this approach presents limitations in both performance and effectiveness. First, the timeout-based crash detection mechanism introduces delays, which impact on the fuzzing throughput. Second, the lack of visibility into the target system's internal state prevents the fuzz driver from leveraging feedback mechanisms like edge coverage, potentially reducing its effectiveness in uncovering vulnerabilities that may be hidden deeper in the firmware logic.

\bigbreak
\noindent
\subsection{Fuzzing Depth}
To evaluate the effectiveness of \toolname{} against Linux-based firmware compared to traditional network fuzzing, we measured code coverage (edge and basic blocks) achieved during fuzzing and assessed the ability of both setups to rediscover known vulnerabilities.
For all evaluation targets in both fuzzing setups, we used a valid PCAP as fuzzing seed, which contains HTTP requests that exercise a known vulnerable functionality of the target firmware. 
We used the same crafted PCAPs for both \toolname{} and the \textit{baseline} setups. 

\hypersetup{
  colorlinks = false,
}

For the Tenda AC15 firmware, we targeted the endpoint ''\url{goform/fast_setting_wifi_set}'', which is used for WiFi setup. This endpoint has a known vulnerability in the ''\texttt{ssid}'' field of the HTTP request, which can be exploited to trigger a buffer overflow (CVE-2018-16333) in the router's web server. 

For the TEW-651BR firmware, we targeted the \texttt{get\_set.ccp} endpoint, and we specifically fuzzed the ''\texttt{ccp\_act}'' parameter passed through the body of the HTTP request. The manipulation of this parameter can lead to a memory corruption vulnerability (CVE-2019-11400), potentially having impact on confidentiality, integrity and availability. 

Finally, for the DCS-932L firmware, we targeted the \texttt{wireless.htm} endpoint and the WEPencryption field, leading to a stack-based buffer overflow (CVE-2019-10999) in the camera's web server. This overflow can potentially allow a remotely authenticated attacker to execute arbitrary code by providing a long string in the ''\texttt{WEPEncryption}'' parameter when requesting wireless.htm. 

In summary, \toolname{} intercepts and selectively fuzzes bytes in the buffer received by the \texttt{recvfrom()} system call while processing incoming HTTP requests. In contrast, the \textit{Baseline} setup mutates bytes in the PCAP before sending (corrupted) requests to the firmware network interface, similarly to existing fuzzing tools for HTTP such as Burp Suite and OWASP ZAP.

\bigbreak
\noindent
\textbf{\textit{Coverage Analysis}} We measured the edge and block coverage achieved by \toolname{} and the \textit{Baseline} approach on Linux-based firmware. Figure \ref{fig:coverage_firmware} illustrates the coverage growth curves over time, demonstrating that \toolname{} consistently outperforms the black-box network fuzzing approach in two out of three cases and performs comparably in the third one.

\begin{figure*}[t]
  \includegraphics[width=\textwidth]{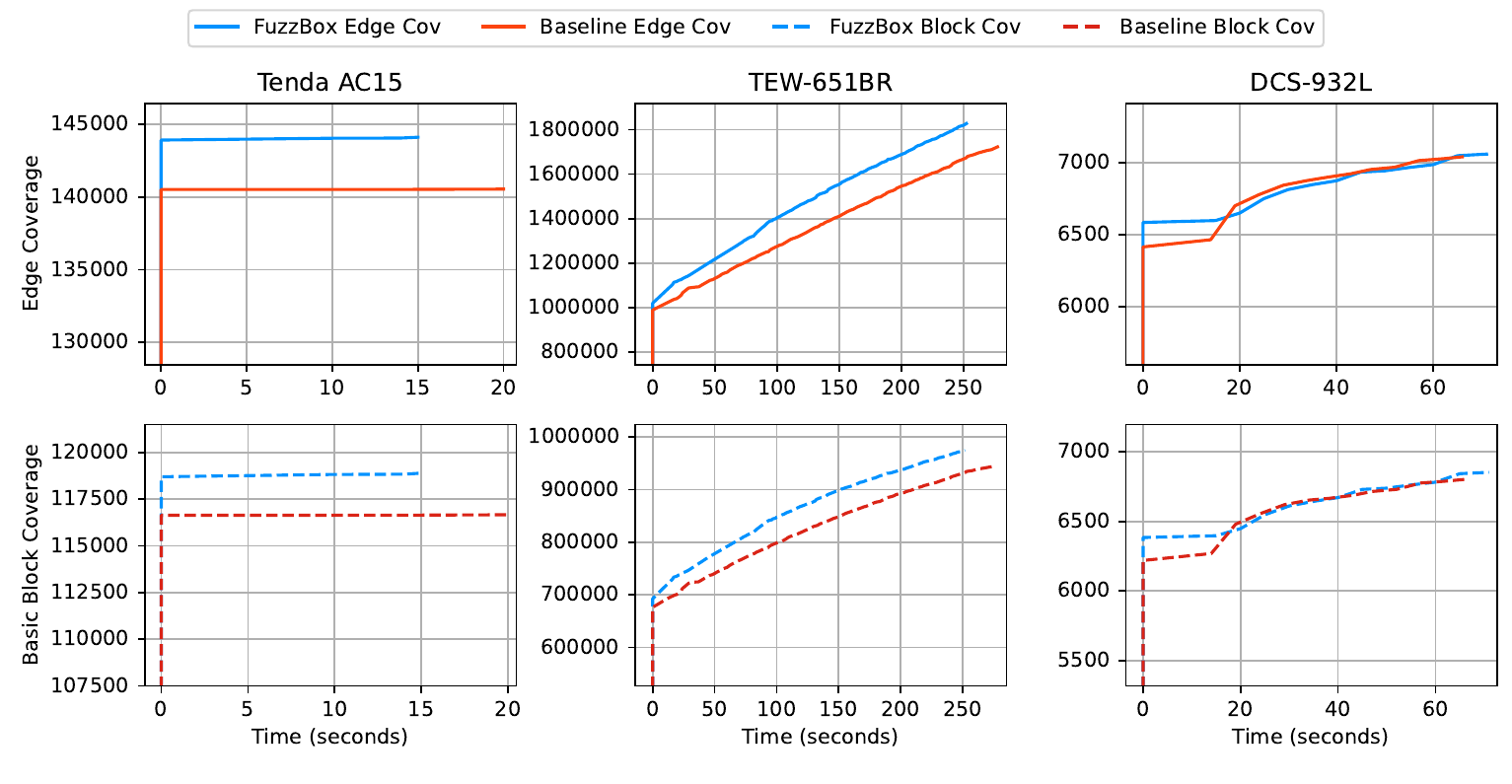}
  \caption{Coverage growth curves over time for Linux-based firmware targets.}
  \label{fig:coverage_firmware}
\end{figure*}

\bigbreak
\noindent
\textbf{\textit{Bug Analysis.}} In our experiment, we successfully discovered three known vulnerabilities in the target firmware. Leveraging grey-box fuzzing, fewer or same iterations were required to explore deep paths within the target and effectively trigger these vulnerabilities. Table \ref{tab:bug_discovery_iot} summarizes findings on bug discovery, reporting the required iterations to crash, as well as the fuzzing seed, comparing the two approaches.

\verbdef\seedA|ssid=seed|
\verbdef\seedB|ccp_act=aaaaaa|
\verbdef\seedC|WEPEncryption=seed|

\verbdef\stringaA|seed=???^z|
\verbdef\stringaB|seed=??^Y?????????????????????????^\^\^|
\verbdef\stringaC|seed=?aaaaa?Aa?aaaa?aaaaAa^\a?aaaaaa|
\verbdef\stringaD|seed=aaaaaaaaa?aaaaaaaajaaa??aaaaaa??|
\verbdef\stringaE|seed=s^Q^Q^Q^Q^Q^Q^Q^Q^QQ?^Q^Q^Q^Q^Q^Q^Q^Q|
\verbdef\stringaF|seed=s^Q^Q^Q^Q^Q^Q^Q^Q^QQ^Q^Q^Q^Q^Q^Q^Q^Q|

\verbdef\num|#|

\begin{table}[h]
  \centering
  \scriptsize
  \caption{Vulnerabilities in \toolname{} and \textit{Baseline} for Linux-based firmware.}
  \label{tab:bug_discovery_iot}
  
  \begin{tabular}{p{0.15\linewidth} p{0.2\linewidth} p{0.3\linewidth} p{0.05\linewidth} p{0.12\linewidth}}
    \toprule
    \textbf{Target} & \textbf{CVE} & \textbf{Seed} & \textbf{Iter.} & \textbf{Approach} \\
    \midrule

    \multirow{2}{*}{\textit{TENDA AC15}} & \multirow{2}{*}{CVE-2018-16333} & \multirow{2}{*}{\seedA}  & \cellcolor{green!30} 25 & \cellcolor{green!30} \toolname{}\\
               &                 & & \cellcolor{red!30} 39 & \cellcolor{red!30} \textit{Baseline}\\

    \arrayrulecolor{customgrey}\hline\arrayrulecolor{black}
    
    \multirow{2}{*}{\textit{TEW-651BR}}  &  \multirow{2}{*}{CVE-2019-11400} & \multirow{2}{*}{\seedB}   & \cellcolor{green!30} 114 & \cellcolor{green!30} \toolname{}\\
                &                &  & \cellcolor{red!30} 145 &\cellcolor{red!30} \textit{Baseline}\\

    \arrayrulecolor{customgrey}\hline\arrayrulecolor{black}
    
    \multirow{2}{*}{\textit{DCS-932L}}  &  \multirow{2}{*}{CVE-2019-10999)} & \multirow{2}{*}{\seedC} &  \cellcolor{green!30} 13  & \cellcolor{green!30} \toolname{}\\
                &                &                 & \cellcolor{green!30} 13 &\cellcolor{green!30} \textit{Baseline}\\

    \bottomrule
  \end{tabular}
\end{table}

\Definition{
The evaluation highlights \toolname{}'s efficiency and effectiveness in identifying critical security weaknesses in Linux-based IoT firmware, extending its applicability beyond its focus on industrial systems. In this context, unlike traditional network fuzzing, which submits corrupted HTTP requests externally, \toolname{} employs on-the-fly corruption of internal IPC interfaces, potentially enabling deeper control flow manipulation. This innovative approach allows it to uncover vulnerabilities that may remain unreachable through network-level fuzzing alone. By achieving higher code coverage and efficiently rediscovering known vulnerabilities, \toolname{} demonstrates its portability to diverse embedded environments. While primarily designed for industrial systems, this use case underscores its adaptability to other embedded systems, such as IoT firmware, further validating its versatility.
}

\section{Discussion}
\label{sec:discussion}
Despite positive results obtained by using the proposed architecture against MILS-based applications, we briefly discuss limitations and avenues of further improvement.

\bigbreak
\noindent
\textbf{\textit{Emulation Accuracy.}} Our approach relies on emulation for interception, fuzzing, and feedback. However, there are potential limitations when the target system can not be fully emulated. This might occur due to lacking support for specific peripherals or lower accuracy in replicating real-world interactions, which can impact fuzzing effectiveness in finding security flaws. This could result in false positives or negatives, and limitations in code coverage. To tackle these challenges, we can explore solutions like using the QEMU extensions for automated rehosting of unsupported peripherals \cite{liu2021firmguide, johnson2021jetset, fasano2021sok}. Incorporating Hardware-In-The-Loop (HIL) techniques \cite{diaz2021enabling} is another option while keeping our approach for integrating fuzzing with emulation. Despite these limitations, emulation remains a popular choice for assessing embedded systems \cite{srivastava2019firmfuzz, gao2020fuzz, zheng2022efficient}.

\bigbreak
\noindent
\textbf{\textit{Symbol Table Dependency.}} \toolname{} relies on the symbol table to get information about the target function. This table is usually included in the binary during development if debug information is enabled. MILS binaries in ELF format support this. Other binary formats like COFF, PE, and Mach-O, also include it. In production builds, debug information might be stripped to reduce binary size. While it is not the primary purpose of \toolname{}, in these use cases, reverse engineering tools like Fuzzable \cite{fuzzable} can be used automatically discover fuzzable target functions through static analysis of the binary. Alternatively, advanced techniques such as probabilistic naming of functions can be applied to stripped binaries \cite{patrick2020probabilistic}.

\bigbreak
\noindent
\textbf{\textit{Symbol Table Dependency.}} 
\toolname{} relies on the availability of symbol information to locate and intercept target functions. This is feasible when working with development binaries that include debug symbols, which is often the case for MILS kernels and applications during integration testing. However, production-grade binaries frequently strip symbol tables to minimize size and hinder reverse engineering. In these cases, users must identify target functions without direct symbol references.

Several techniques can support this process. Static binary analysis tools like Fuzzable \cite{fuzzable} or Angr can be used to infer potential fuzzable functions based on control-flow patterns or binary signatures. Other research efforts have proposed probabilistic function naming and similarity-based matching to recover semantic information from stripped binaries \cite{patrick2020probabilistic}. Although symbol inference is not the primary goal of \toolname{}, integrating such capabilities remains a promising extension for increasing applicability to closed-source and production binaries.

\bigbreak
\noindent
\textbf{\textit{Security Mitigations.}} Certain security measures at the binary level, such as Control Flow Integrity (CFI), Data Execution Prevention (DEP), and stack canaries can co-exist without significantly hindering the \textit{FuzzBox} approach. The primary concern is Address Space Layout Randomization (ASLR), which randomizes the memory layout of the target and introduces unpredictability in system call addresses. While ASLR could be a hurdle, the strategic use of memory forensics tools, such as Volatility \cite{Volatility}, \cite{orbinato2023laccolith}, emerges as a promising solution for dynamically identifying syscall addresses during runtime. Nevertheless, protections like ASLR are typically enabled for the production release of the target, but not during the testing phase.

\bigbreak
\noindent
\textbf{\textit{Portability.}} While this paper focuses on MILS-based applications, the \toolname{} design principles can extend to various embedded systems, such as IoT firmware and industrial control system applications. For example, the tool can adapt to intercept and fuzz network packet handlers in these systems. Its modular design, with configurable components for profiling, injection, and coverage collection, facilitates straightforward reconfiguration for diverse target systems and architectures. Although \toolname{} inherently supports such scenarios by design, future efforts should focus on evaluating its effectiveness across different targets and architectures. 

\bigbreak
\noindent
\textbf{\textit{Automatic Configuration.}} 
While \toolname{} currently requires manual effort to configure target functions, parameters to fuzz, and architecture-specific settings (see Section~\ref{subsec:configuration}), future work could automate these steps. First, target functions in binaries can be identified using heuristics (e.g., \cite{fuzzable}, \cite{patrick2020probabilistic}) based on naming conventions, interface signatures, and known APIs (e.g., POSIX or MILS primitives). Dynamic analysis, such as frequent I/O interactions and common call patterns, can refine these candidates. In addition, profiles of known kernels and libraries can be shared for reuse. Second, static analysis tools, such as Ghidra, can help infer pointer arguments and associated size fields to automate parameter selection. Third, although architecture-specific calling conventions must be respected, \toolname{} already includes profiles for PowerPC, ARM, and MIPS. Future work could extend support to additional architectures, and analysis of binary metadata (e.g., ELF headers) can guide automatic selection of the architecture profile.

\section{Conclusion}
\label{sec:conclusion}
We presented \toolname{}, a tool for fuzzing binary-only industrial targets. The core idea of \toolname{} is to integrate fuzzing with the dynamic binary translation performed to virtualize targets. This approach enables the tracking of function invocations, the injection of fuzz inputs, the detection of failures, and the profiling of coverage in a hardware-independent, build toolchain-independent, and non-intrusive manner. Therefore, the tool supports the application of grey-box fuzzing in the context of industrial systems, by reducing the effort needed to setup a fuzzing workflow.

\backmatter

\section*{Declarations}

\bmhead{Funding}
Not applicable

\bmhead{Ethics approval and consent to participate}
Not applicable

\bmhead{Availability of data and material}
Not applicable

\bmhead{Code availability}
The code use to support the findings of this study is available in the \url{https://github.com/dessertlab/FuzzBox} repository.

\bmhead{Acknowledgements}
This project has been partially supported by the GENIO project (CUP B69J23005770005) funded by MIMIT, ''Accordi per l'Innovazione'' program.

\bibliography{bibliography}


\begin{thebibliography}{69}
\ifx \bisbn   \undefined \def \bisbn  #1{ISBN #1}\fi
\ifx \binits  \undefined \def \binits#1{#1}\fi
\ifx \bauthor  \undefined \def \bauthor#1{#1}\fi
\ifx \batitle  \undefined \def \batitle#1{#1}\fi
\ifx \bjtitle  \undefined \def \bjtitle#1{#1}\fi
\ifx \bvolume  \undefined \def \bvolume#1{\textbf{#1}}\fi
\ifx \byear  \undefined \def \byear#1{#1}\fi
\ifx \bissue  \undefined \def \bissue#1{#1}\fi
\ifx \bfpage  \undefined \def \bfpage#1{#1}\fi
\ifx \blpage  \undefined \def \blpage #1{#1}\fi
\ifx \burl  \undefined \def \burl#1{\textsf{#1}}\fi
\ifx \doiurl  \undefined \def \doiurl#1{\url{https://doi.org/#1}}\fi
\ifx \betal  \undefined \def \betal{\textit{et al.}}\fi
\ifx \binstitute  \undefined \def \binstitute#1{#1}\fi
\ifx \binstitutionaled  \undefined \def \binstitutionaled#1{#1}\fi
\ifx \bctitle  \undefined \def \bctitle#1{#1}\fi
\ifx \beditor  \undefined \def \beditor#1{#1}\fi
\ifx \bpublisher  \undefined \def \bpublisher#1{#1}\fi
\ifx \bbtitle  \undefined \def \bbtitle#1{#1}\fi
\ifx \bedition  \undefined \def \bedition#1{#1}\fi
\ifx \bseriesno  \undefined \def \bseriesno#1{#1}\fi
\ifx \blocation  \undefined \def \blocation#1{#1}\fi
\ifx \bsertitle  \undefined \def \bsertitle#1{#1}\fi
\ifx \bsnm \undefined \def \bsnm#1{#1}\fi
\ifx \bsuffix \undefined \def \bsuffix#1{#1}\fi
\ifx \bparticle \undefined \def \bparticle#1{#1}\fi
\ifx \barticle \undefined \def \barticle#1{#1}\fi
\bibcommenthead
\ifx \bconfdate \undefined \def \bconfdate #1{#1}\fi
\ifx \botherref \undefined \def \botherref #1{#1}\fi
\ifx \url \undefined \def \url#1{\textsf{#1}}\fi
\ifx \bchapter \undefined \def \bchapter#1{#1}\fi
\ifx \bbook \undefined \def \bbook#1{#1}\fi
\ifx \bcomment \undefined \def \bcomment#1{#1}\fi
\ifx \oauthor \undefined \def \oauthor#1{#1}\fi
\ifx \citeauthoryear \undefined \def \citeauthoryear#1{#1}\fi
\ifx \endbibitem  \undefined \def \endbibitem {}\fi
\ifx \bconflocation  \undefined \def \bconflocation#1{#1}\fi
\ifx \arxivurl  \undefined \def \arxivurl#1{\textsf{#1}}\fi
\csname PreBibitemsHook\endcsname

%%% 1
\bibitem[\protect\citeauthoryear{Zeller et~al.}{2019}]{zeller2019fuzzing}
\begin{botherref}
\oauthor{\bsnm{Zeller}, \binits{A.}},
\oauthor{\bsnm{Gopinath}, \binits{R.}},
\oauthor{\bsnm{B{\"o}hme}, \binits{M.}},
\oauthor{\bsnm{Fraser}, \binits{G.}},
\oauthor{\bsnm{Holler}, \binits{C.}}:
The fuzzing book.
CISPA+ Saarland University
(2019)
\end{botherref}
\endbibitem

%%% 2
\bibitem[\protect\citeauthoryear{Man{\`e}s et~al.}{2019}]{manes2018fuzzing}
\begin{botherref}
\oauthor{\bsnm{Man{\`e}s}, \binits{V.J.}},
\oauthor{\bsnm{Han}, \binits{H.}},
\oauthor{\bsnm{Han}, \binits{C.}},
\oauthor{\bsnm{Cha}, \binits{S.K.}},
\oauthor{\bsnm{Egele}, \binits{M.}},
\oauthor{\bsnm{Schwartz}, \binits{E.J.}},
\oauthor{\bsnm{Woo}, \binits{M.}}:
The art, science, and engineering of fuzzing: A survey.
IEEE Transactions on Software Engineering
(2019)
\end{botherref}
\endbibitem

%%% 3
\bibitem[\protect\citeauthoryear{Eisele et~al.}{2022}]{eisele2022embedded}
\begin{barticle}
\bauthor{\bsnm{Eisele}, \binits{M.}},
\bauthor{\bsnm{Maugeri}, \binits{M.}},
\bauthor{\bsnm{Shriwas}, \binits{R.}},
\bauthor{\bsnm{Huth}, \binits{C.}},
\bauthor{\bsnm{Bella}, \binits{G.}}:
\batitle{Embedded fuzzing: a review of challenges, tools, and solutions}.
\bjtitle{Cybersecurity}
\bvolume{5}(\bissue{1}),
\bfpage{18}
(\byear{2022})
\end{barticle}
\endbibitem

%%% 4
\bibitem[\protect\citeauthoryear{Yun et~al.}{2022}]{yun2022fuzzing}
\begin{botherref}
\oauthor{\bsnm{Yun}, \binits{J.}},
\oauthor{\bsnm{Rustamov}, \binits{F.}},
\oauthor{\bsnm{Kim}, \binits{J.}},
\oauthor{\bsnm{Shin}, \binits{Y.}}:
{Fuzzing of Embedded Systems: A Survey}.
ACM Computing Surveys
(2022)
\end{botherref}
\endbibitem

%%% 5
\bibitem[\protect\citeauthoryear{}{2023}]{llvm}
\begin{botherref}
{The LLVM Compiler Infrastructure}.
\url{https://llvm.org/}
(2023)
\end{botherref}
\endbibitem

%%% 6
\bibitem[\protect\citeauthoryear{}{2023}]{gcc}
\begin{botherref}
{GNU GCC}.
\url{https://gcc.gnu.org/}
(2023)
\end{botherref}
\endbibitem

%%% 7
\bibitem[\protect\citeauthoryear{}{2018}]{afl-dyninst}
\begin{botherref}
{talos-vulndev - AFL-DynInst}.
\url{https://github.com/talos-vulndev/afl-dyninst}
(2018)
\end{botherref}
\endbibitem

%%% 8
\bibitem[\protect\citeauthoryear{Dinesh et~al.}{2020}]{dinesh2020retrowrite}
\begin{bchapter}
\bauthor{\bsnm{Dinesh}, \binits{S.}},
\bauthor{\bsnm{Burow}, \binits{N.}},
\bauthor{\bsnm{Xu}, \binits{D.}},
\bauthor{\bsnm{Payer}, \binits{M.}}:
\bctitle{{Retrowrite: Statically instrumenting cots binaries for fuzzing and sanitization}}.
In: \bbtitle{2020 IEEE Symposium on Security and Privacy (SP)}
(\byear{2020}).
\bcomment{IEEE}
\end{bchapter}
\endbibitem

%%% 9
\bibitem[\protect\citeauthoryear{}{2018}]{AFL-DynamoRIO}
\begin{botherref}
{Heuse, Marc - AFL-DynamoRIO}.
\url{https://github.com/vanhauser-thc/afl-dynamorio}
(2018)
\end{botherref}
\endbibitem

%%% 10
\bibitem[\protect\citeauthoryear{}{2020}]{AFL-PIN}
\begin{botherref}
{Heuse, Marc - AFL-PIN}.
\url{https://github.com/vanhauser-thc/afl-pin}
(2020)
\end{botherref}
\endbibitem

%%% 11
\bibitem[\protect\citeauthoryear{Chen et~al.}{2019}]{chen2019ptrix}
\begin{bchapter}
\bauthor{\bsnm{Chen}, \binits{Y.}},
\bauthor{\bsnm{Mu}, \binits{D.}},
\bauthor{\bsnm{Xu}, \binits{J.}},
\bauthor{\bsnm{Sun}, \binits{Z.}},
\bauthor{\bsnm{Shen}, \binits{W.}},
\bauthor{\bsnm{Xing}, \binits{X.}},
\bauthor{\bsnm{Lu}, \binits{L.}},
\bauthor{\bsnm{Mao}, \binits{B.}}:
\bctitle{{Ptrix: Efficient hardware-assisted fuzzing for cots binary}}.
In: \bbtitle{Proceedings of the 2019 ACM Asia Conference on Computer and Communications Security}
(\byear{2019})
\end{bchapter}
\endbibitem

%%% 12
\bibitem[\protect\citeauthoryear{Schumilo et~al.}{2017}]{schumilo2017kafl}
\begin{bchapter}
\bauthor{\bsnm{Schumilo}, \binits{S.}},
\bauthor{\bsnm{Aschermann}, \binits{C.}},
\bauthor{\bsnm{Gawlik}, \binits{R.}},
\bauthor{\bsnm{Schinzel}, \binits{S.}},
\bauthor{\bsnm{Holz}, \binits{T.}}:
\bctitle{$\{$kAFL$\}$:$\{$Hardware-Assisted$\}$ feedback fuzzing for $\{$OS$\}$ kernels}.
In: \bbtitle{26th USENIX Security Symposium (USENIX Security 17)}
(\byear{2017})
\end{bchapter}
\endbibitem

%%% 13
\bibitem[\protect\citeauthoryear{}{2016}]{honggfuzz}
\begin{botherref}
{Robert, Swiecki - Honggfuzz}.
\url{http://code.google.com/p/honggfuzz}
(2016)
\end{botherref}
\endbibitem

%%% 14
\bibitem[\protect\citeauthoryear{Eisele et~al.}{2023}]{eisele2023fuzzing}
\begin{bchapter}
\bauthor{\bsnm{Eisele}, \binits{M.}},
\bauthor{\bsnm{Ebert}, \binits{D.}},
\bauthor{\bsnm{Huth}, \binits{C.}},
\bauthor{\bsnm{Zeller}, \binits{A.}}:
\bctitle{Fuzzing embedded systems using debug interfaces}.
In: \bbtitle{Proceedings of ACM SIGSOFT International Symposium on Software Testing and Analysis (ISSTA 2023).}
(\byear{2023})
\end{bchapter}
\endbibitem

%%% 15
\bibitem[\protect\citeauthoryear{Li et~al.}{2022}]{li2022muafl}
\begin{bchapter}
\bauthor{\bsnm{Li}, \binits{W.}},
\bauthor{\bsnm{Shi}, \binits{J.}},
\bauthor{\bsnm{Li}, \binits{F.}},
\bauthor{\bsnm{Lin}, \binits{J.}},
\bauthor{\bsnm{Wang}, \binits{W.}},
\bauthor{\bsnm{Guan}, \binits{L.}}:
\bctitle{$\mu$afl: non-intrusive feedback-driven fuzzing for microcontroller firmware}.
In: \bbtitle{Proceedings of the 44th International Conference on Software Engineering}
(\byear{2022})
\end{bchapter}
\endbibitem

%%% 16
\bibitem[\protect\citeauthoryear{Neugass et~al.}{1991}]{neugass1991vxworks}
\begin{bchapter}
\bauthor{\bsnm{Neugass}, \binits{H.}},
\bauthor{\bsnm{Espin}, \binits{G.}},
\bauthor{\bsnm{Nunoe}, \binits{H.}},
\bauthor{\bsnm{Thomas}, \binits{R.}},
\bauthor{\bsnm{Wilner}, \binits{D.}}:
\bctitle{Vxworks: an interactive development environment and real-time kernel for gmicro}.
In: \bbtitle{Eighth TRON Project Symposium}
(\byear{1991}).
\bcomment{IEEE Computer Society}
\end{bchapter}
\endbibitem

%%% 17
\bibitem[\protect\citeauthoryear{Alves-Foss et~al.}{2006}]{alves2006mils}
\begin{botherref}
\oauthor{\bsnm{Alves-Foss}, \binits{J.}},
\oauthor{\bsnm{Oman}, \binits{P.W.}},
\oauthor{\bsnm{Taylor}, \binits{C.}},
\oauthor{\bsnm{Harrison}, \binits{W.S.}}:
{The MILS architecture for high-assurance embedded systems}.
International journal of embedded systems
(2006)
\end{botherref}
\endbibitem

%%% 18
\bibitem[\protect\citeauthoryear{Netkachova et~al.}{2015}]{netkachova2015security}
\begin{bchapter}
\bauthor{\bsnm{Netkachova}, \binits{K.}},
\bauthor{\bsnm{M{\"u}ller}, \binits{K.}},
\bauthor{\bsnm{Paulitsch}, \binits{M.}},
\bauthor{\bsnm{Bloomfield}, \binits{R.}}:
\bctitle{Security-informed safety case approach to analysing mils systems}.
In: \bbtitle{International Workshop on MILS: Architecture and Assurance for Secure Systems}
(\byear{2015})
\end{bchapter}
\endbibitem

%%% 19
\bibitem[\protect\citeauthoryear{{Airlines Electronic Engineering Committee}}{2005}]{airlines2005arinc}
\begin{botherref}
\oauthor{\bsnm{{Airlines Electronic Engineering Committee}}}:
{ARINC 811: Commercial aircraft information security concepts of operation and process framework}.
Aeronautical Radio, Inc
(2005)
\end{botherref}
\endbibitem

%%% 20
\bibitem[\protect\citeauthoryear{Yang et~al.}{2009}]{yang2009inter}
\begin{bchapter}
\bauthor{\bsnm{Yang}, \binits{X.}},
\bauthor{\bsnm{Lei}, \binits{J.}},
\bauthor{\bsnm{Xiong}, \binits{G.-z.}}:
\bctitle{{Inter-partition Information Flow Control for High-Assurance Embedded Systems}}.
In: \bbtitle{2009 WRI World Congress on Computer Science and Information Engineering}
(\byear{2009}).
\bcomment{IEEE}
\end{bchapter}
\endbibitem

%%% 21
\bibitem[\protect\citeauthoryear{}{2023}]{windriver_workbench}
\begin{botherref}
{Wind River} - {VxWorks Workbench}.
\url{https://docs.windriver.com/bundle/Workbench_4_Getting_Started_OpenVersion_23_03/page/elx1502803804955.html}
(2023)
\end{botherref}
\endbibitem

%%% 22
\bibitem[\protect\citeauthoryear{Ispoglou et~al.}{2020}]{ispoglou2020fuzzgen}
\begin{bchapter}
\bauthor{\bsnm{Ispoglou}, \binits{K.}},
\bauthor{\bsnm{Austin}, \binits{D.}},
\bauthor{\bsnm{Mohan}, \binits{V.}},
\bauthor{\bsnm{Payer}, \binits{M.}}:
\bctitle{$\{$FuzzGen$\}$: Automatic fuzzer generation}.
In: \bbtitle{29th USENIX Security Symposium 20}
(\byear{2020})
\end{bchapter}
\endbibitem

%%% 23
\bibitem[\protect\citeauthoryear{Feng et~al.}{2021}]{feng2021snipuzz}
\begin{bchapter}
\bauthor{\bsnm{Feng}, \binits{X.}},
\bauthor{\bsnm{Sun}, \binits{R.}},
\bauthor{\bsnm{Zhu}, \binits{X.}},
\bauthor{\bsnm{Xue}, \binits{M.}},
\bauthor{\bsnm{Wen}, \binits{S.}},
\bauthor{\bsnm{Liu}, \binits{D.}},
\bauthor{\bsnm{Nepal}, \binits{S.}},
\bauthor{\bsnm{Xiang}, \binits{Y.}}:
\bctitle{Snipuzz: Black-box fuzzing of iot firmware via message snippet inference}.
In: \bbtitle{Proceedings of the 2021 ACM SIGSAC Conference on Computer and Communications Security}
(\byear{2021})
\end{bchapter}
\endbibitem

%%% 24
\bibitem[\protect\citeauthoryear{}{}]{vxworks_board}
\begin{botherref}
{Wind RiverVxWorks MILS 3.0 BSP for WR SBC8548}.
\url{https://bsp.windriver.com/bsps/1573}
\end{botherref}
\endbibitem

%%% 25
\bibitem[\protect\citeauthoryear{Du et~al.}{2022}]{du2022afliot}
\begin{barticle}
\bauthor{\bsnm{Du}, \binits{X.}},
\bauthor{\bsnm{Chen}, \binits{A.}},
\bauthor{\bsnm{He}, \binits{B.}},
\bauthor{\bsnm{Chen}, \binits{H.}},
\bauthor{\bsnm{Zhang}, \binits{F.}},
\bauthor{\bsnm{Chen}, \binits{Y.}}:
\batitle{{AflIot: Fuzzing on linux-based IoT device with binary-level instrumentation}}.
\bjtitle{Computers \& Security}
\bvolume{122},
\bfpage{102889}
(\byear{2022})
\end{barticle}
\endbibitem

%%% 26
\bibitem[\protect\citeauthoryear{}{2017}]{afl}
\begin{botherref}
{{M. Zalewski - American fuzzy lop}}, author = {}.
\url{http://lcamtuf.coredump.cx/afl/}
(2017)
\end{botherref}
\endbibitem

%%% 27
\bibitem[\protect\citeauthoryear{Fioraldi et~al.}{2020}]{fioraldi2020afl++}
\begin{bchapter}
\bauthor{\bsnm{Fioraldi}, \binits{A.}},
\bauthor{\bsnm{Maier}, \binits{D.}},
\bauthor{\bsnm{Ei{\ss}feldt}, \binits{H.}},
\bauthor{\bsnm{Heuse}, \binits{M.}}:
\bctitle{$\{$AFL++$\}$: Combining incremental steps of fuzzing research}.
In: \bbtitle{14th USENIX Workshop on Offensive Technologies (WOOT 20)}
(\byear{2020})
\end{bchapter}
\endbibitem

%%% 28
\bibitem[\protect\citeauthoryear{Maier et~al.}{2019}]{maier2019unicorefuzz}
\begin{bchapter}
\bauthor{\bsnm{Maier}, \binits{D.}},
\bauthor{\bsnm{Radtke}, \binits{B.}},
\bauthor{\bsnm{Harren}, \binits{B.}}:
\bctitle{Unicorefuzz: On the viability of emulation for kernelspace fuzzing}.
In: \bbtitle{13th USENIX Workshop on Offensive Technologies (WOOT 19)}
(\byear{2019})
\end{bchapter}
\endbibitem

%%% 29
\bibitem[\protect\citeauthoryear{}{2015}]{vyukov2015syzkaller}
\begin{botherref}
google - {Syzkaller}.
\url{https://github.com/google/syzkaller}
(2015)
\end{botherref}
\endbibitem

%%% 30
\bibitem[\protect\citeauthoryear{Shen et~al.}{2022}]{shen2022tardis}
\begin{botherref}
\oauthor{\bsnm{Shen}, \binits{Y.}},
\oauthor{\bsnm{Xu}, \binits{Y.}},
\oauthor{\bsnm{Sun}, \binits{H.}},
\oauthor{\bsnm{Liu}, \binits{J.}},
\oauthor{\bsnm{Xu}, \binits{Z.}},
\oauthor{\bsnm{Cui}, \binits{A.}},
\oauthor{\bsnm{Shi}, \binits{H.}},
\oauthor{\bsnm{Jiang}, \binits{Y.}}:
{Tardis: Coverage-Guided Embedded Operating System Fuzzing}.
IEEE Transactions on Computer-Aided Design of Integrated Circuits and Systems
(2022)
\end{botherref}
\endbibitem

%%% 31
\bibitem[\protect\citeauthoryear{Bellard}{2005}]{bellard2005qemu}
\begin{bchapter}
\bauthor{\bsnm{Bellard}, \binits{F.}}:
\bctitle{Qemu, a fast and portable dynamic translator.}
In: \bbtitle{USENIX Annual Technical Conference, FREENIX Track}
(\byear{2005}).
\bcomment{Califor-nia, USA}
\end{bchapter}
\endbibitem

%%% 32
\bibitem[\protect\citeauthoryear{Isovic and Fohler}{2000}]{isovic2000efficient}
\begin{bchapter}
\bauthor{\bsnm{Isovic}, \binits{D.}},
\bauthor{\bsnm{Fohler}, \binits{G.}}:
\bctitle{Efficient scheduling of sporadic, aperiodic, and periodic tasks with complex constraints}.
In: \bbtitle{Proceedings 21st IEEE Real-Time Systems Symposium}
(\byear{2000}).
\bcomment{IEEE}
\end{bchapter}
\endbibitem

%%% 33
\bibitem[\protect\citeauthoryear{}{}]{fuzzable}
\begin{botherref}
{Fuzzable - Framework for Automating Fuzzable Target Discovery with Static Analysis}.
\url{https://github.com/ex0dus-0x/fuzzable}
\end{botherref}
\endbibitem

%%% 34
\bibitem[\protect\citeauthoryear{}{2023}]{powerPCconvention}
\begin{botherref}
{IBM - PowerPC register usage conventions}.
\url{https://www.ibm.com/docs/en/aix/7.2?topic=overview-register-usage-conventions}
(2023)
\end{botherref}
\endbibitem

%%% 35
\bibitem[\protect\citeauthoryear{Babi{\'c} et~al.}{2019}]{babic2019fudge}
\begin{bchapter}
\bauthor{\bsnm{Babi{\'c}}, \binits{D.}},
\bauthor{\bsnm{Bucur}, \binits{S.}},
\bauthor{\bsnm{Chen}, \binits{Y.}},
\bauthor{\bsnm{Ivan{\v{c}}i{\'c}}, \binits{F.}},
\bauthor{\bsnm{King}, \binits{T.}},
\bauthor{\bsnm{Kusano}, \binits{M.}},
\bauthor{\bsnm{Lemieux}, \binits{C.}},
\bauthor{\bsnm{Szekeres}, \binits{L.}},
\bauthor{\bsnm{Wang}, \binits{W.}}:
\bctitle{{FUDGE}: Fuzz driver generation at scale}.
In: \bbtitle{Proceedings of the 2019 27th ACM Joint Meeting on European Software Engineering Conference and Symposium on the Foundations of Software Engineering}
(\byear{2019})
\end{bchapter}
\endbibitem

%%% 36
\bibitem[\protect\citeauthoryear{Jeong et~al.}{2023}]{jeong2023utopia}
\begin{bchapter}
\bauthor{\bsnm{Jeong}, \binits{B.}},
\bauthor{\bsnm{Jang}, \binits{J.}},
\bauthor{\bsnm{Yi}, \binits{H.}},
\bauthor{\bsnm{Moon}, \binits{J.}},
\bauthor{\bsnm{Kim}, \binits{J.}},
\bauthor{\bsnm{Jeon}, \binits{I.}},
\bauthor{\bsnm{Kim}, \binits{T.}},
\bauthor{\bsnm{Shim}, \binits{W.}},
\bauthor{\bsnm{Hwang}, \binits{Y.H.}}:
\bctitle{{UTopia}: Automatic generation of fuzz driver using unit tests}.
In: \bbtitle{2023 IEEE Symposium on Security and Privacy (SP)}
(\byear{2023}).
\bcomment{IEEE}
\end{bchapter}
\endbibitem

%%% 37
\bibitem[\protect\citeauthoryear{}{2023}]{qemu_architecture_support}
\begin{botherref}
{QEMU Architectures Support}.
\url{https://wiki.qemu.org/Documentation/Platforms}
(2023)
\end{botherref}
\endbibitem

%%% 38
\bibitem[\protect\citeauthoryear{}{2019}]{adacore_qemu}
\begin{botherref}
{QEMU AdaCore}.
\url{https://github.com/AdaCore/qemu/blob/qemu-stable-4.0.0/hw/ppc/p2010rdb.c}
(2019)
\end{botherref}
\endbibitem

%%% 39
\bibitem[\protect\citeauthoryear{Fioraldi et~al.}{2022}]{fioraldi2022libafl}
\begin{bchapter}
\bauthor{\bsnm{Fioraldi}, \binits{A.}},
\bauthor{\bsnm{Maier}, \binits{D.C.}},
\bauthor{\bsnm{Zhang}, \binits{D.}},
\bauthor{\bsnm{Balzarotti}, \binits{D.}}:
\bctitle{Libafl: A framework to build modular and reusable fuzzers}.
In: \bbtitle{Proceedings of the 2022 ACM SIGSAC Conference on Computer and Communications Security}
(\byear{2022})
\end{bchapter}
\endbibitem

%%% 40
\bibitem[\protect\citeauthoryear{}{2023}]{TCG_plugin}
\begin{botherref}
{QEMU TCG Plugins}.
\url{https://github.com/qemu/qemu/blob/master/docs/devel/tcg-plugins.rst}
(2023)
\end{botherref}
\endbibitem

%%% 41
\bibitem[\protect\citeauthoryear{}{2022}]{JsonParser}
\begin{botherref}
{json-parser}.
\url{https://github.com/json-parser/json-parser}
(2022)
\end{botherref}
\endbibitem

%%% 42
\bibitem[\protect\citeauthoryear{}{2020}]{SendMail}
\begin{botherref}
sendmail.
\url{https://github.com/mykter/afl-training/tree/main/challenges/sendmail/1305}
(2020)
\end{botherref}
\endbibitem

%%% 43
\bibitem[\protect\citeauthoryear{}{2022}]{TinyExpr}
\begin{botherref}
{TinyExpr}.
\url{https://github.com/codeplea/tinyexpr}
(2022)
\end{botherref}
\endbibitem

%%% 44
\bibitem[\protect\citeauthoryear{Rushby}{1981}]{rushby1981design}
\begin{barticle}
\bauthor{\bsnm{Rushby}, \binits{J.M.}}:
\batitle{Design and verification of secure systems}.
\bjtitle{ACM SIGOPS Operating Systems Review}
\bvolume{15}(\bissue{5}),
\bfpage{12}--\blpage{21}
(\byear{1981})
\end{barticle}
\endbibitem

%%% 45
\bibitem[\protect\citeauthoryear{Klees et~al.}{2018}]{klees2018evaluating}
\begin{bchapter}
\bauthor{\bsnm{Klees}, \binits{G.}},
\bauthor{\bsnm{Ruef}, \binits{A.}},
\bauthor{\bsnm{Cooper}, \binits{B.}},
\bauthor{\bsnm{Wei}, \binits{S.}},
\bauthor{\bsnm{Hicks}, \binits{M.}}:
\bctitle{Evaluating fuzz testing}.
In: \bbtitle{Proceedings of the 2018 ACM SIGSAC Conference on Computer and Communications Security}
(\byear{2018})
\end{bchapter}
\endbibitem

%%% 46
\bibitem[\protect\citeauthoryear{Dolan-Gavitt et~al.}{2016}]{dolan2016lava}
\begin{bchapter}
\bauthor{\bsnm{Dolan-Gavitt}, \binits{B.}},
\bauthor{\bsnm{Hulin}, \binits{P.}},
\bauthor{\bsnm{Kirda}, \binits{E.}},
\bauthor{\bsnm{Leek}, \binits{T.}},
\bauthor{\bsnm{Mambretti}, \binits{A.}},
\bauthor{\bsnm{Robertson}, \binits{W.}},
\bauthor{\bsnm{Ulrich}, \binits{F.}},
\bauthor{\bsnm{Whelan}, \binits{R.}}:
\bctitle{Lava: Large-scale automated vulnerability addition}.
In: \bbtitle{2016 IEEE Symposium on Security and Privacy (SP)}
(\byear{2016}).
\bcomment{IEEE}
\end{bchapter}
\endbibitem

%%% 47
\bibitem[\protect\citeauthoryear{Tsoutsos and Maniatakos}{2018}]{anatomy2018}
\begin{barticle}
\bauthor{\bsnm{Tsoutsos}, \binits{N.G.}},
\bauthor{\bsnm{Maniatakos}, \binits{M.}}:
\batitle{{Anatomy of Memory Corruption Attacks and Mitigations in Embedded Systems}}.
\bjtitle{IEEE Embedded Systems Letters}
\bvolume{10}(\bissue{3}),
\bfpage{95}--\blpage{98}
(\byear{2018})
\end{barticle}
\endbibitem

%%% 48
\bibitem[\protect\citeauthoryear{Islam and Muzahid}{2013}]{Characterizing2013}
\begin{bchapter}
\bauthor{\bsnm{Islam}, \binits{M.M.}},
\bauthor{\bsnm{Muzahid}, \binits{A.}}:
\bctitle{Characterizing real world bugs causing sequential consistency violations}.
In: \bbtitle{Proceedings of the 5th USENIX Conference on Hot Topics in Parallelism}.
\bsertitle{HotPar'13},
p. \bfpage{8}.
\bpublisher{USENIX Association},
\blocation{USA}
(\byear{2013})
\end{bchapter}
\endbibitem

%%% 49
\bibitem[\protect\citeauthoryear{Geng et~al.}{2020}]{geng2020empirical}
\begin{botherref}
\oauthor{\bsnm{Geng}, \binits{S.}},
\oauthor{\bsnm{Li}, \binits{Y.}},
\oauthor{\bsnm{Du}, \binits{Y.}},
\oauthor{\bsnm{Xu}, \binits{J.}},
\oauthor{\bsnm{Liu}, \binits{Y.}},
\oauthor{\bsnm{Mao}, \binits{B.}}:
{An empirical study on benchmarks of artificial software vulnerabilities}.
arXiv preprint arXiv:2003.09561
(2020)
\end{botherref}
\endbibitem

%%% 50
\bibitem[\protect\citeauthoryear{Kim et~al.}{2020}]{kim2020firmae}
\begin{bchapter}
\bauthor{\bsnm{Kim}, \binits{M.}},
\bauthor{\bsnm{Kim}, \binits{D.}},
\bauthor{\bsnm{Kim}, \binits{E.}},
\bauthor{\bsnm{Kim}, \binits{S.}},
\bauthor{\bsnm{Jang}, \binits{Y.}},
\bauthor{\bsnm{Kim}, \binits{Y.}}:
\bctitle{Firmae: Towards large-scale emulation of iot firmware for dynamic analysis}.
In: \bbtitle{Proceedings of the 36th Annual Computer Security Applications Conference},
pp. \bfpage{733}--\blpage{745}
(\byear{2020})
\end{bchapter}
\endbibitem

%%% 51
\bibitem[\protect\citeauthoryear{Kim et~al.}{2021}]{kim2021firm}
\begin{barticle}
\bauthor{\bsnm{Kim}, \binits{J.}},
\bauthor{\bsnm{Yu}, \binits{J.}},
\bauthor{\bsnm{Kim}, \binits{H.}},
\bauthor{\bsnm{Rustamov}, \binits{F.}},
\bauthor{\bsnm{Yun}, \binits{J.}}:
\batitle{Firm-cov: high-coverage greybox fuzzing for iot firmware via optimized process emulation}.
\bjtitle{IEEE Access}
\bvolume{9},
\bfpage{101627}--\blpage{101642}
(\byear{2021})
\end{barticle}
\endbibitem

%%% 52
\bibitem[\protect\citeauthoryear{Zheng et~al.}{2019}]{zheng2019firm}
\begin{bchapter}
\bauthor{\bsnm{Zheng}, \binits{Y.}},
\bauthor{\bsnm{Davanian}, \binits{A.}},
\bauthor{\bsnm{Yin}, \binits{H.}},
\bauthor{\bsnm{Song}, \binits{C.}},
\bauthor{\bsnm{Zhu}, \binits{H.}},
\bauthor{\bsnm{Sun}, \binits{L.}}:
\bctitle{$\{$FIRM-AFL$\}$:$\{$High-Throughput$\}$ greybox fuzzing of $\{$IoT$\}$ firmware via augmented process emulation}.
In: \bbtitle{28th USENIX Security Symposium (USENIX Security 19)},
pp. \bfpage{1099}--\blpage{1114}
(\byear{2019})
\end{bchapter}
\endbibitem

%%% 53
\bibitem[\protect\citeauthoryear{}{2024}]{tenda_ac15}
\begin{botherref}
{Tenda - AC15 Firmware}.
\url{https://www.tendacn.com/it/product/download/a15.html}
(2024)
\end{botherref}
\endbibitem

%%% 54
\bibitem[\protect\citeauthoryear{}{2024}]{tew_firmware}
\begin{botherref}
{Trendnet - TEW-651BR Firmware}.
\url{https://www.trendnet.com/support/support-detail.asp?prod=190_TEW-651BR}
(2024)
\end{botherref}
\endbibitem

%%% 55
\bibitem[\protect\citeauthoryear{}{2024}]{DSC_firmware}
\begin{botherref}
{Dlink - DCS-932L Firmware}.
\url{https://support.dlink.com.au/download/download.aspx?product=DCS-932L}
(2024)
\end{botherref}
\endbibitem

%%% 56
\bibitem[\protect\citeauthoryear{Pham et~al.}{2020}]{pham2020aflnet}
\begin{bchapter}
\bauthor{\bsnm{Pham}, \binits{V.-T.}},
\bauthor{\bsnm{B{\"o}hme}, \binits{M.}},
\bauthor{\bsnm{Roychoudhury}, \binits{A.}}:
\bctitle{Aflnet: a greybox fuzzer for network protocols}.
In: \bbtitle{2020 IEEE 13th International Conference on Software Testing, Validation and Verification (ICST)}
(\byear{2020}).
\bcomment{IEEE}
\end{bchapter}
\endbibitem

%%% 57
\bibitem[\protect\citeauthoryear{}{2024}]{burpsuite}
\begin{botherref}
{Port Swigger - BurpSuite}.
\url{https://portswigger.net/burp}
(2024)
\end{botherref}
\endbibitem

%%% 58
\bibitem[\protect\citeauthoryear{Owasp}{2024}]{owasp_zap}
\begin{botherref}
\oauthor{\bsnm{Owasp}}:
ZAP.
\url{https://www.zaproxy.org/}
(2024)
\end{botherref}
\endbibitem

%%% 59
\bibitem[\protect\citeauthoryear{Chen et~al.}{2018}]{chen2018iotfuzzer}
\begin{bchapter}
\bauthor{\bsnm{Chen}, \binits{J.}},
\bauthor{\bsnm{Diao}, \binits{W.}},
\bauthor{\bsnm{Zhao}, \binits{Q.}},
\bauthor{\bsnm{Zuo}, \binits{C.}},
\bauthor{\bsnm{Lin}, \binits{Z.}},
\bauthor{\bsnm{Wang}, \binits{X.}},
\bauthor{\bsnm{Lau}, \binits{W.C.}},
\bauthor{\bsnm{Sun}, \binits{M.}},
\bauthor{\bsnm{Yang}, \binits{R.}},
\bauthor{\bsnm{Zhang}, \binits{K.}}:
\bctitle{Iotfuzzer: Discovering memory corruptions in iot through app-based fuzzing.}
In: \bbtitle{NDSS}
(\byear{2018})
\end{bchapter}
\endbibitem

%%% 60
\bibitem[\protect\citeauthoryear{Liu et~al.}{2021}]{liu2021firmguide}
\begin{bchapter}
\bauthor{\bsnm{Liu}, \binits{Q.}},
\bauthor{\bsnm{Zhang}, \binits{C.}},
\bauthor{\bsnm{Ma}, \binits{L.}},
\bauthor{\bsnm{Jiang}, \binits{M.}},
\bauthor{\bsnm{Zhou}, \binits{Y.}},
\bauthor{\bsnm{Wu}, \binits{L.}},
\bauthor{\bsnm{Shen}, \binits{W.}},
\bauthor{\bsnm{Luo}, \binits{X.}},
\bauthor{\bsnm{Liu}, \binits{Y.}},
\bauthor{\bsnm{Ren}, \binits{K.}}:
\bctitle{Firmguide: Boosting the capability of rehosting embedded linux kernels through model-guided kernel execution}.
In: \bbtitle{2021 36th IEEE/ACM International Conference on Automated Software Engineering (ASE)}
(\byear{2021}).
\bcomment{IEEE}
\end{bchapter}
\endbibitem

%%% 61
\bibitem[\protect\citeauthoryear{Johnson et~al.}{2021}]{johnson2021jetset}
\begin{bchapter}
\bauthor{\bsnm{Johnson}, \binits{E.}},
\bauthor{\bsnm{Bland}, \binits{M.}},
\bauthor{\bsnm{Zhu}, \binits{Y.}},
\bauthor{\bsnm{Mason}, \binits{J.}},
\bauthor{\bsnm{Checkoway}, \binits{S.}},
\bauthor{\bsnm{Savage}, \binits{S.}},
\bauthor{\bsnm{Levchenko}, \binits{K.}}:
\bctitle{Jetset: Targeted firmware rehosting for embedded systems}.
In: \bbtitle{30th USENIX Security Symposium 21}
(\byear{2021})
\end{bchapter}
\endbibitem

%%% 62
\bibitem[\protect\citeauthoryear{Fasano et~al.}{2021}]{fasano2021sok}
\begin{bchapter}
\bauthor{\bsnm{Fasano}, \binits{A.}},
\bauthor{\bsnm{Ballo}, \binits{T.}},
\bauthor{\bsnm{Muench}, \binits{M.}},
\bauthor{\bsnm{Leek}, \binits{T.}},
\bauthor{\bsnm{Bulekov}, \binits{A.}},
\bauthor{\bsnm{Dolan-Gavitt}, \binits{B.}},
\bauthor{\bsnm{Egele}, \binits{M.}},
\bauthor{\bsnm{Francillon}, \binits{A.}},
\bauthor{\bsnm{Lu}, \binits{L.}},
\bauthor{\bsnm{Gregory}, \binits{N.}}, \betal:
\bctitle{Sok: Enabling security analyses of embedded systems via rehosting}.
In: \bbtitle{Proceedings of the 2021 ACM Asia Conference on Computer and Communications Security}
(\byear{2021})
\end{bchapter}
\endbibitem

%%% 63
\bibitem[\protect\citeauthoryear{D{\'\i}az et~al.}{2021}]{diaz2021enabling}
\begin{botherref}
\oauthor{\bsnm{D{\'\i}az}, \binits{E.}},
\oauthor{\bsnm{Mateos}, \binits{R.}},
\oauthor{\bsnm{Bueno}, \binits{E.J.}},
\oauthor{\bsnm{Nieto}, \binits{R.}}:
Enabling parallelized-qemu for hardware/software co-simulation virtual platforms.
Electronics
(2021)
\end{botherref}
\endbibitem

%%% 64
\bibitem[\protect\citeauthoryear{Srivastava et~al.}{2019}]{srivastava2019firmfuzz}
\begin{bchapter}
\bauthor{\bsnm{Srivastava}, \binits{P.}},
\bauthor{\bsnm{Peng}, \binits{H.}},
\bauthor{\bsnm{Li}, \binits{J.}},
\bauthor{\bsnm{Okhravi}, \binits{H.}},
\bauthor{\bsnm{Shrobe}, \binits{H.}},
\bauthor{\bsnm{Payer}, \binits{M.}}:
\bctitle{{Firmfuzz: Automated IoT firmware introspection and analysis}}.
In: \bbtitle{Proceedings of the 2nd International ACM Workshop on Security and Privacy for the Internet-of-Things}
(\byear{2019})
\end{bchapter}
\endbibitem

%%% 65
\bibitem[\protect\citeauthoryear{Gao et~al.}{2020}]{gao2020fuzz}
\begin{botherref}
\oauthor{\bsnm{Gao}, \binits{J.}},
\oauthor{\bsnm{Xu}, \binits{Y.}},
\oauthor{\bsnm{Jiang}, \binits{Y.}},
\oauthor{\bsnm{Liu}, \binits{Z.}},
\oauthor{\bsnm{Chang}, \binits{W.}},
\oauthor{\bsnm{Jiao}, \binits{X.}},
\oauthor{\bsnm{Sun}, \binits{J.}}:
Em-fuzz: Augmented firmware fuzzing via memory checking.
IEEE Transactions on Computer-Aided Design of Integrated Circuits and Systems
(2020)
\end{botherref}
\endbibitem

%%% 66
\bibitem[\protect\citeauthoryear{Zheng et~al.}{2022}]{zheng2022efficient}
\begin{bchapter}
\bauthor{\bsnm{Zheng}, \binits{Y.}},
\bauthor{\bsnm{Li}, \binits{Y.}},
\bauthor{\bsnm{Zhang}, \binits{C.}},
\bauthor{\bsnm{Zhu}, \binits{H.}},
\bauthor{\bsnm{Liu}, \binits{Y.}},
\bauthor{\bsnm{Sun}, \binits{L.}}:
\bctitle{{Efficient greybox fuzzing of applications in Linux-based IoT devices via enhanced user-mode emulation}}.
In: \bbtitle{Proceedings of the 31st ACM SIGSOFT International Symposium on Software Testing and Analysis}
(\byear{2022})
\end{bchapter}
\endbibitem

%%% 67
\bibitem[\protect\citeauthoryear{Patrick-Evans et~al.}{2020}]{patrick2020probabilistic}
\begin{bchapter}
\bauthor{\bsnm{Patrick-Evans}, \binits{J.}},
\bauthor{\bsnm{Cavallaro}, \binits{L.}},
\bauthor{\bsnm{Kinder}, \binits{J.}}:
\bctitle{Probabilistic naming of functions in stripped binaries}.
In: \bbtitle{Annual Computer Security Applications Conference}
(\byear{2020})
\end{bchapter}
\endbibitem

%%% 68
\bibitem[\protect\citeauthoryear{}{}]{Volatility}
\begin{botherref}
{Volatility}.
\url{https://www.volatilityfoundation.org/}
\end{botherref}
\endbibitem

%%% 69
\bibitem[\protect\citeauthoryear{Orbinato et~al.}{2024}]{orbinato2023laccolith}
\begin{botherref}
\oauthor{\bsnm{Orbinato}, \binits{V.}},
\oauthor{\bsnm{Feliciano}, \binits{M.C.}},
\oauthor{\bsnm{Cotroneo}, \binits{D.}},
\oauthor{\bsnm{Natella}, \binits{R.}}:
Laccolith: Hypervisor-based adversary emulation with anti-detection.
IEEE Transactions on Dependable and Secure Computing
(2024)
\end{botherref}
\endbibitem

\end{thebibliography}

\end{document}